\documentclass[prb,nobalancelastpage,showpacs,twocolumn,superscriptaddress,floatfix]{revtex4}

\usepackage[normalem]{ulem}  % package containing \sout command

\usepackage{graphicx}
\usepackage{hyperref} 

\usepackage{amssymb}
\usepackage{amsmath}
\usepackage{eucal}
\usepackage{color}
\usepackage{bm}
\usepackage{dashrule}

\newcommand{\e}{{\rm e}}
\newcommand{\rmd}{{\rm d}}
\newcommand{\rmi}{{\rm i}}

\newcommand{\al}{\alpha}

\newcommand{\kB}{ k_{\rm B} }
\newcommand{\dtraj}{\gamma}
\newcommand{\dtrajdiag}{\gamma_{\rm d}}

\newcommand{\Esr}{E_{\mbox{{\scriptsize s-r}}}}

\newcommand{\ignore}[1]{\relax}

 \definecolor{DarkRed}{rgb}{0.8,0,0}
 \definecolor{DarkBlue}{rgb}{0,0,0.8}
\newcommand{\dRed}[1]{\textcolor{DarkRed}{#1}}

\renewcommand{\sout}[1]{}

\begin{document}

\title{Non-Markovian quantum thermodynamics: laws \& fluctuation theorems}

\author{Robert S.~Whitney}
\affiliation{
Laboratoire de Physique et Mod\'elisation des Milieux Condens\'es (UMR 5493), 
Universit\'e Grenoble Alpes and CNRS, Maison des Magist\`eres, BP 166, 38042 Grenoble, France.}

\date{April 11, 2018}
\begin{abstract}
This work brings together thermodynamics and non-equilibrium quantum theory, by showing that a real-time diagrammatic technique
on the Keldysh contour is an equivalent of stochastic thermodynamics for non-Markovian quantum machines (heat engines, refrigerators, etc).  Symmetries are found between quantum trajectories and their time-reverses on the Keldysh contour, for any interacting quantum system coupled to ideal reservoirs of electrons, phonons or photons.  These lead to quantum fluctuation theorems the same as the well-known classical ones (Jarzynski and Crooks equalities, integral fluctuation theorem, etc), whether the system's dynamics are Markovian or not.  Some of these are also shown to hold for non-factorizable initial states. 
The sequential tunnelling approximation and the cotunnelling approximation are both shown to respect the symmetries that ensure the fluctuation theorems.  For all initial states, energy conservation ensures that the first law of thermodynamics holds on average, while the above symmetries ensures that the
second law of thermodynamics holds on average, even if fluctuations violate it.  
\dRed{\textsf{[ERRATUM added: March 2021]}}
\end{abstract}

% 73.63.-b Electronic transport in mesoscopic systems
% 05.30.-d	Quantum statistical mechanics
% 05.70.Ln, nonequilibrium thermodynamics
% 05.10.Gg	Stochastic analysis methods (Fokker-Planck, Langevin, etc.)
% 72.15.Jf  thermoelectric effects (metals and alloys)
% 84.60.Rb  Thermoelectric energy conversion

\maketitle
%=====================================================

%%%%%%%%%%%%%%%%%%%%%%%%%%%%
\begin{figure}[b]
\includegraphics[width=0.85\columnwidth]{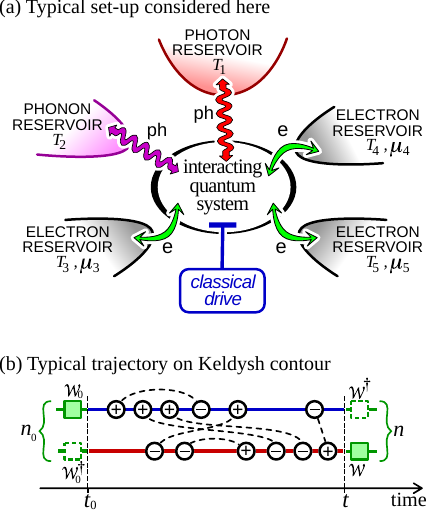}
\caption{\label{Fig:sys+diagram}
(a) This work considers a quantum system coupled to any number of 
electron reservoirs with chemical potentials and temperatures $\{\mu_\alpha,T_\alpha\}$,
and photon or phonon reservoirs at temperatures $\{T_\alpha\}$.
(b) A typical double Keldysh  trajectory, $\dtraj$, in which the horizontal lines represent the evolution of the system state, while the dashed-lines indicate transitions within the system due to the coupling to one of the reservoirs.}
\end{figure}
%%%%%%%%%%%%%%%%%%%%%%%%%%

\vskip -1mm\noindent
\dRed{\textsf{ERRATUM (14 March 2021). The published version of this article contained a stupid error in the condition for the validity of the Crooks equaton. This error is corrected here by the red text in section~\ref{Sect:Crooks}}}.

\section{Introduction}
The laws of thermodynamics were derived for macroscopic machines,
where entropy-reducing fluctuations (e.g.\ a gas spontaneously drifting into one corner of its container) are so rare that they have been referred to as {\it thermodynamic miracles}. \cite{Thermdynamic-Miracles}  
In microscopic systems on short timescales, these ``miracles'' are rather common, 
and we now know they obey fluctuation theorems.\cite{Derrida-CRAS2007,Searles-Review-2008,Campisi-review2011,qu-thermo-review-Anders,qu-thermo-review-Millen}
There is a unifying theory of such theorems in classical systems called
{\it stochastic thermodynamics},\cite{Seifert-PRL2005,Schmiedl-Seifert2007}
reviewed in Refs.~[\onlinecite{Seifert-review2012,van-Broeck-review2015,Seifert-PRL2016,our-review-2017}]. 
It gives the 
Jarzynski \cite{Jarzynski}, Evans-Searles \cite{Evans-Searles1994} and Crooks \cite{Crooks1999,Tasaki} equalities in the relevant limits.
It was used to show \cite{Seifert-PRL2005,Schmiedl-Seifert2007} that {\it any} classical system with Markovian dynamics
obeys
\begin{eqnarray}
\big\langle e^{-\Delta S_{\rm tot}} \big\rangle =1\ ,
\label{Eq:non-equil_partition_identity}
\end{eqnarray}
where $\Delta S_{\rm tot}$ is the total entropy change of the system and reservoirs,\cite{footnote:definitionS} and the average is over all possible thermal fluctuations.\cite{Footnote:units}
This has become known as the {\it integral fluctuation theorem},\cite{Seifert-PRL2005,Schmiedl-Seifert2007,Seifert-review2012,van-Broeck-review2015} even if a similar identity had appeared under the name
{\it non-equilibrium partition identity} earlier.\cite{Yamada-Kawasaki1967,Morriss1985,Carberry2005}
Eq.~(\ref{Eq:non-equil_partition_identity}) tells us that  the second law of thermodynamics is obeyed on average, $\langle \Delta S_{\rm tot} \rangle \geq 0$.  Yet Eq.~(\ref{Eq:non-equil_partition_identity}) also tells us that fluctuations with $\Delta S_{\rm tot} < 0$ {\it must} occur (even if rarely), otherwise $\big\langle \e^{-\Delta S_{\rm tot}} \big\rangle$ would be less than one.

At the same time, there is great interest in the thermodynamics of nanoscale machines, particularly those which convert heat into electricity, or use electricity to perform refrigeration.  Such machines are definitely not macroscopic, so we can expect them to exhibit fluctuations similar to those described above.  However most of them also exhibit quantum effects that are not captured by classical theory of stochastic thermodynamics.
Many operate in the steady-state, such as the quantum-dot heat-engines experimentally realized in Refs.~\cite{Roche15,Hartmann15,Thierschmann15}, or other mesoscopic systems which exhibit thermoelectric effects \cite{our-review-2017,CRAS}, while others involve pumping cycles \cite{Juergens2013}.  The general case of such a machine is sketched in Fig.~\ref{Fig:sys+diagram}a.

This work shows that a diagrammatic technique on the Keldysh contour --- real-time transport theory
\cite{Schoeller-Schon1994,Konig96,Konig97,Schoeller-review1997} ---
provides an equivalent of stochastic thermodynamics 
for any quantum system coupled to reservoirs 
(Fig.~\ref{Fig:sys+diagram}a) 
whether that system's dynamics are Markovian or not.
It makes the connection between the contribution of a double-trajectory, $\dtraj$, on the Keldysh contour
and the contribution of its time-reverse, $\overline\dtraj$  (Fig.~\ref{Fig:rotate}a).   
This is enough to show that such systems respect the same fluctuation theorems as classical 
Markovian systems, and so obey the second law of thermodynamics on average.
For the second law, our proof goes beyond
those for Markovian quantum systems \cite{Kosloff-review2013}, 
those for systems with mean-field interactions,\cite{Nenciu2007,2012w-2ndlaw} and 
Keldysh treatments for non-interacting systems (quadratic Hamiltonians) \cite{Sanchez-2ndlaw2014,Esposito-2ndlaw,Bruch-2016} or adiabatic driving \cite{Sanchez-2ndlaw-2016}
based on the Keldysh techniques reviewed in Ref.~[\onlinecite{Kamenev-book}].
This connection between fluctuation theorems \cite{Campisi-review2011,qu-thermo-review-Anders,qu-thermo-review-Millen} and a non-equilibrium quantum theory for transport through interacting systems,\cite{Schoeller-Schon1994,Konig96,Konig97,Schoeller-review1997,Leijnse2008,Schoeller2009,Wegewijs2014,Sothmann2014,Schulenborg2016}
provides a powerful tool for modelling energy production and refrigeration at the nanoscale.
In this context, significant currents and power outputs require significant system-reservoir coupling.
However, only systems in the weak coupling limit have Markovian dynamics,\cite{Davies74,Davies76}.
Thus there is great interest in improving the power output of experimental set-ups like the quantum dot heat-engines in 
Refs.~\cite{Roche15,Hartmann15,Thierschmann15} by taking them to stronger coupling,
where their dynamics will be non-Markovian systems.

Previous proofs of fluctuation theorems in non-Markovian quantum systems 
exist,\cite{Campisi-review2011} but rely on treating the system and reservoirs together as a single isolated quantum system.
This is elegant, but is not amenable to calculating a given machine's power or efficiency, except in the rare cases where the full Hamiltonian (system plus reservoirs) is exactly soluble. 
It gives no indication of what approximations allow calculations of this power or efficiency, without an unphysical violation of fluctuation theorems and of the second law of thermodynamics.
This work finds a microscopic symmetry which underlies the fluctuation theorems, 
beyond the Markovian quantum systems considered in Ref.~[\onlinecite{Maxime+Alexia2016}].
This enables one to identify a family of approximations that allow tractable calculations of machine power and efficiency, with no risk of violating the second law or fluctuation theorems. 
%These approximations 
%can model phenomena such as Coulomb blockade,  Kondo effect, etc \cite{Schoeller-Schon1994,Konig96,Konig97,Schoeller-review1997,Leijnse2008,Schoeller2009,Wegewijs2014,Sothmann2014,Schulenborg2016}.  

\subsection{Overview of this work}

The central observation of this work is the result connecting trajectories on the Keldysh contour 
in a system, to time-reversed trajectories in a time-reversed system, given in section ~\ref{Sect:Traj-TR} and shown schematically in Fig.~\ref{Fig:rotate}.  The sections leading up to section~\ref{Sect:TR-derivation} set the scene, with section~\ref{Sect:firstlaw} making the observation that such trajectories obey the first law of thermodynamics on average.
Then section~\ref{Sect:TR-derivation} itself provides a derivation of the result connecting trajectories to their time-reverse.

The rest of this work then uses this result in deriving various fluctuation theorems.
Section~\ref{Sect:fluct} uses it to derive various fluctuation theorems in various situations, such as the Jarzynski equality and the Crooks equation.  In particular, it shows that 
the integral fluctuation theorem in Eq.~(\ref{Eq:non-equil_partition_identity}) holds for any system which starts in a product state with the reservoirs,  thereby showing that any such system obeys the second law of thermodynamics on average.
Section~\ref{Sect:nonfactorized} provides similar proofs for situations where the system and reservoir start in a non-factorizable initial state.
Finally, section~\ref{Sect:approx} discusses approximations that respect 
the result in section~\ref{Sect:Traj-TR}, and thereby will not violate any of the fluctuation theorems in 
sections~\ref{Sect:fluct} and \ref{Sect:nonfactorized}, and so will always satisfy the second law of thermodynamics
on average.

\subsection{A comment on the system-reservoir coupling}
\label{Sect:comment-sys-res}
The recent literature on Keldysh for quantum thermodynamics \cite{Sanchez-2ndlaw2014,Esposito-2ndlaw,Bruch-2016,Sanchez-2ndlaw-2016} 
has strongly debated the role of the average energy stored in the system-reservoir coupling, $\big\langle\Esr\big\rangle$, in models of adiabatic pumping or 
in which interactions are absent.
The initial claim was that $\big\langle\Esr\big\rangle$ should be separated into two equal parts, with one part being assigned to system and the other part to the reservoir, 
however Ref.~[\onlinecite{Bruch-2016}] argued that this was mainly a matter of calculational convenience.

In light of this debate, it is worth mentioning the role of $\big\langle\Esr\big\rangle$ in the diagrammatic approach used here,
whose differences from the approach in Refs.~[\onlinecite{Sanchez-2ndlaw2014,Esposito-2ndlaw,Bruch-2016,Sanchez-2ndlaw-2016}],
are described in section~\ref{Sect:H} below. 
Firstly, $\big\langle\Esr\big\rangle$ appears in the first law, but does not appear in the second law or the integral fluctuation theorem, since they involve entropy rather than energy. Secondly, it is {\it not} convenient to assign any part of $\big\langle\Esr\big\rangle$ to the reservoirs for the following reason.
In the cases considered here, each reservoir is in a state that is ``simple'', that is to say  it is in local equilibrium, which is completely described by two parameters; temperature and electrochemical potential. 
However, the system state is not ``simple'' in this sense, because it is typically far from equilibrium (due to the action of multiple reservoirs and/or driving), and requires more than just these two parameters to describe it.  This means that the system-reservoir coupling is also not ``simple''.  Hence, it is unhelpful to associate any part of the system-reservoir coupling with the reservoir state, because one then loses the simplicity of the latter.
In contrast the energy in the system-reservoir coupling always appears together with the system energy (see section~\ref{Sect:firstlaw}), and as neither contribution is ``simple'' in the above sense, there is no disadvantage with making the choice to  combine the two into a single {\it effective} internal system energy.  
That said, this article keeps the energy in the system-reservoir coupling separate from the system energy throughout, to avoid ambiguity.

\section{Hamiltonian}
\label{Sect:H}

This work considers 
a small quantum system with the Hamiltonian, $\hat{H}_{\rm sys}(t)$, which may include a time-dependent driving and interactions between the particles in the system.  This system (shown at the centre of Fig.~\ref{Fig:sys+diagram}a) acts as a machine changing the heat and work in the reservoirs that surround it.
Each term in $\hat{H}_{\rm sys}(t)$ contains one creation operator for a system electronic state, $\hat{d}_i^\dagger$, for every annihilation operator, $\hat{d}_j$.
This system is coupled to multiple reservoirs of non-interacting fermions (electrons) via couplings  $\hat{V}_{\rm el}^{(\alpha)}(t)$,
or non-interacting bosons (photons or phonons) via couplings $\hat{V}_{\rm ph}^{(\alpha)}(t)$.   
This article uses the word ``set-up'' to refer to the system and reservoirs together; 
the total Hamiltonian of this set-up is
\begin{eqnarray}
\hat{H}_{\rm tot}(t) &=& \hat{H}_{\rm sys}(t) 
+\sum_{\alpha \in {\rm el}}\!
\left[ \hat{V}_{\rm el}^{(\alpha)}(t) +\hat{H}_{\rm el}^{(\alpha)} \right]
\nonumber \\
& & \qquad\qquad\qquad\ +\sum_{\alpha \in {\rm ph}} \!\!
\left[ \hat{V}_{\rm ph}^{(\alpha)}(t) +\hat{H}_{\rm ph}^{(\alpha)} \right] \! . \qquad
\label{Eq:H_tot}
\end{eqnarray}
The sums are over electron (el) and photon/phonon (ph) reservoirs.
For el reservoirs, 
\begin{eqnarray}
\hat{H}_{\rm el}^{(\alpha)}  = \sum_k \,E_{\alpha k}\, \hat{c}_{\alpha k}^\dagger \,\hat{c}_{\alpha k},
\end{eqnarray}
for reservoir $\alpha$'s state $k$ with energy, creation and annihilation operators 
 $E_{\alpha k}$, $\hat{c}_{\alpha k}^\dagger$ and $\hat{c}_{\alpha k}$.
The tunnel coupling,
\begin{eqnarray}
\hat{V}_{\rm el}^{(\alpha)}(t) =  \sum_{k} \,\big( 
 \hat{V}^+_{\alpha k}(t)\, \hat{c}_{\alpha k} + \hat{V}^-_{\alpha k}(t)\, \hat{c}_{\alpha k}^\dagger  \big),
  \label{Eq:coupling-el}
 \end{eqnarray}
where $\hat{V}^-_{\alpha k}(t)$ and $\hat{V}^+_{\alpha k}(t)$ contain
only system operators, and may be time-dependent. 
The change in the system state when an electron is added from reservoir $\alpha$'s state $k$
is given by $\hat{V}^+_{\alpha k}$.  
The reverse process is given by $\hat{V}^-_{\alpha k}=\big[\hat{V}^+_{\alpha k}\big]^\dagger$.
The simplest case has
$\hat{V}^+_{\alpha k} =\sum_i A_{ik}^{(\alpha)}\hat{d}_i^\dagger$, however if 
the coupling depends on the system state, then  $\hat{V}^+_{\alpha k}$ contains extra factors of $\hat{d}_j^\dagger\hat{d}_{j'}$.
For bosonic reservoirs, one replaces the fermionic operators 
$\hat{c}_{\alpha k}^\dagger$ and $\hat{c}_{\alpha k}$ by bosonic ones.  
The simplest case has 
$\hat{V}^+_{\alpha k} =\sum_{ij} A_{ijk}^{(\alpha)}\hat{d}_i^\dagger\hat{d}_j$,
meaning the system goes from $j$ to $i$ when a boson is absorbed from reservoir $\alpha$'s state $k$.   
%In general $\hat{V}^+_{\alpha k}$ can have multiple terms, in which each term has one system creation operator for every annihilation operator.  

The first step to using the real-time transport theory\cite{Schoeller-Schon1994,Konig96,Konig97,Schoeller-review1997} is to write all system operators as $N \times N$ matrices acting on the basis of $N$ many-body system states, see e.g.\ appendix C of Ref.~[\onlinecite{our-review-2017}]. 
We go to an interaction representation (indicated by calligraphic symbols), where system operators evolve under a matrix 
\begin{eqnarray}
{\cal U}_{\rm sys}(\tau,t_0) = T \exp\big[ -\rmi \int_{t_0}^\tau {H}_{\rm sys}(t) \rmd t \big],
\end{eqnarray} 
with $T$ indicating time-ordering. Hence,
\begin{align}
{\cal V}^\pm_{\alpha k}(\tau) \ =\ {\cal U}_{\rm sys}^\dagger (\tau;t_0) \ {V}^\pm_{\alpha k}(\tau)\ 
{\cal U}_{\rm sys} (\tau;t_0) .
\label{Eq:V-interaction-representation}
\end{align}
Reservoir operators evolve under $H_{\rm el/ph}^{(\al)}$, 
so we have 
\begin{eqnarray}
\hat{c}_{\alpha k}^\dagger(\tau) &=& \e^{\rmi E_k (\tau-t_0)} \hat{c}_{\alpha k}^\dagger, 
\nonumber \\
\hat{c}_{\alpha k}(\tau) &=& \e^{-\rmi E_k (\tau-t_0)} \hat{c}_{\alpha k}.
\label{Eq:c-interaction-representation}
\end{eqnarray}

The initial condition (at time $t_0$) is an arbitrary system state in a product state 
with the reservoirs.  Each reservoir $\alpha$ is in its local equilibrium with temperature $T_\alpha$ and chemical potential $\mu_\alpha$ ($\mu_\alpha=0$ for reservoirs of photons or phonons).
We treat $H_{\rm sys}$ exactly, and keep the reservoir's effect on the system finite, in the limit of vanishing
reservoir level-spacing.  This requires taking the system's coupling to each reservoir mode to zero, as the density of such modes goes to infinity, so this coupling can be treated at lowest-order (second-order).\cite{Caldiera-Leggett-1983a,Caldiera-Leggett-1983b,Leggett-et-al,Schoeller-review1997}
None the less, the system may interact with any number reservoir modes at one time (all orders of cotunnelling events), and these interactions do not commute.  Upon tracing out the reservoirs, the resulting system dynamics are highly non-Markovian.
Thus its dynamics are not described by Markovian master equations (Lindblad equations),
whose thermodynamics have already been well studied.\cite{Kosloff-review2013}
Thess dynamics are represented in terms of a Keldysh double-trajectory, as in
Fig.~\ref{Fig:sys+diagram}b, where each second-order interaction with a given reservoir mode is 
represented by a pair of interactions joined by a dashed line.

For readers familiar with the Keldysh methods reviewed in Kamenev's textbook \cite{Kamenev-book} and used in Refs.~[\onlinecite{Sanchez-2ndlaw2014,Esposito-2ndlaw,Bruch-2016,Sanchez-2ndlaw-2016}],
we note that the method used here is different at the level of what is treated as a perturbation.
In  Kamevev's textbook, the Hamiltonian is written in the single particle basis; in this basis the Hamiltonian is quadratic in the absence of interactions between particles, and so is exactly soluble. One then uses increasingly sophisticated perturbative techniques to include the interaction terms such as electron-electron interactions (which are quartic in the single-particle operators).  In contrast, this work uses a diagrammatic method on the Keldysh contour referred to as the {\it real-time transport theory}\cite{Schoeller-Schon1994,Konig96,Konig97,Schoeller-review1997}, which takes a different starting point; it starts in the many-body basis for the system Hamiltonian (the fact that it is in real time is not particularly important). 
In this basis, the  physics of the system alone is trivial (including all interaction effects); however the
system-reservoir couplings take forms that are too complicated to treat exactly. Thus, one has moved the difficulty from the interaction terms to the system-reservoir coupling terms.  This is why this coupling must be treated as a perturbation, for which one sums up classes of irreducible diagrams within some suitable approximation scheme.  

%===========
\section{Assumption of no Maxwell demons in reservoirs}
\label{Sect:no-demons}

The equations of classical and quantum physics are reversible.
For example,  if all degrees of freedom in a quantum system were easy to observe and extract work from, the fact that the full wavefunction of the system and reservoirs undergoes unitary evolution means no (von Neumann) entropy is ever produced.  In both classical and quantum physics, entropy production emerges from a physically motivated assumption about which degrees of freedom are easy to observe and extract work from, and which are not.  Typically this assumption separates everything into macroscopic and microscopic dynamics, where macroscopic dynamics are easy to observe and extract work from, while the microscopic dynamics are inaccessible.
All works on thermodynamics make some sort of assumption of this type, explicitly or implicitly. 

The assumption at the basis of this work is presented here,  for compactness it is referred to as the ``assumption of no Maxwell demons in reservoirs''.  It requires that the system operates without knowing microscopic details of the reservoirs, beyond those encoded in the system-reservoir interaction in Eq.~(\ref{Eq:H_tot}).  
For example, this disallows Maxwell's ``observant and neat-fingered'' demons \cite{Maxwell} (which are usually just circuitry built by physicists) which measure individual reservoir states, and then feedback this information by making a change in the time-dependence of $\hat{H}_{\rm sys}(t)$ or $\hat{V}^{(\alpha)}(t)$  in Eq.~(\ref{Eq:H_tot}) which is conditional on the result of the measurement. 
Just as in classical mechanics, assuming no Maxwell demons is crucial in the emergence of the second law from the underlying theory.
This assumption makes all classical correlations and quantum entanglement between system and reservoirs at the end of the evolution irrelevant, since the system cannot extract work from them. 
Hence, one can trace out the reservoirs when calculating system entropy, and vice versa. Further, even though the system pushes certain reservoir modes out of equilibrium, it is assumed
that this information is inaccessible, so no more work can be extracted from the reservoir than if it were in
a thermal state with the same energy.

Superficially, one might thing this work has  nothing to say about experimental implementations of Maxwell demons in quantum systems, similar to Refs.~[\onlinecite{Pekola-Demon,Cottet-Demon}].  
However, in those cases where the demon is completely mechanical (made of some finite number of degrees of freedom coupled to reservoirs with or without 
time-dependent driving), we can include these degrees of freedom in the system Hamiltonian $\hat{H}_{\rm sys}$, and all results in this work apply.

\section{Trajectories}
\label{Sect:trajectories}
Consider a trajectory $\dtraj$ on the Keldysh contour,
whose upper-line goes from system's many-body state $i_0$ at time $t_0$  to $i$ at time $t$, and whose lower line goes from $j_0$ to $j$ (see examples in Fig.~\ref{Fig:rotate}a).
Matrix elements for transitions are time ordered on the upper line and reverse-time ordered 
on the lower line.
Each transition (each dashed-line in $\dtraj$) has a weight determined by whether it is $D_{0\pm}$, $D_{1\pm}$ or $D_{2\pm}$ in Fig.~\ref{Fig:diagram-elements-2nd-order} (see below). 
Real transitions correspond to $D_{1+}$ in Fig.~\ref{Fig:diagram-elements-2nd-order}
and virtual transitions to $D_{0+}$ and $D_{2+}$.
The trajectory's weight, $P(\gamma)$, is the product of all of these factors of $D_{a\pm}$, 
multiplied by a factor of $-1$ for each crossing of dashed-lines.\cite{Schoeller-review1997}
The probability to go from one system state to another in time $t$, is simply the sum of the weights of all trajectories between those states.

The dashed-lines have the following weights,
\begin{eqnarray}
\left[D^{\alpha k}_{0+}\right]^{i'_{m},i'_{n}}_{i_m,i_n}
\! &=& \!
-\left[{\cal V}^-_{\alpha k}(t_m)\right]^{i'_{m}}_{i_m}
\left[{\cal V}^+_{\alpha k}(t_n)\right]^{i'_{n}}_{i_{n}} 
f^+_{\alpha k} \,\e^{\rmi \Phi_k^{mn}}, \ \ \ \
\label{Eq:D0+}
\\
\left[D^{\alpha k}_{1+}\right]^{j_m,i'_n}_{j'_m,i_n}
\! &=&\!  
\left[{\cal V}^-_{\alpha k}(t_m)\right]^{j_{m}}_{j'_{m}} 
\left[{\cal V}^+_{\alpha k}(t_n) \right]^{i'_{n}}_{ i_{n}} 
f^+_{\alpha k} \,\e^{ \rmi \Phi_k^{mn}} , \ 
\label{Eq:D1+}
\\
%\end{eqnarray}
%\begin{eqnarray}
\left[D^{\alpha k}_{2+}\right]^{j_n,j_m}_{j'_n,j'_m}
\! &=& \! 
-\left[{\cal V}^-_{\alpha k}(t_n)\right]^{j_{n}}_{j'_{n}}
\left[{\cal V}^+_{\alpha k}(t_m)\right]^{j_{m}}_{j'_{m}}  
f^+_{\alpha k}\, \e^{\rmi \Phi_k^{mn}}, \ \ 
\label{Eq:D2+}
\end{eqnarray}
where $[{\cal V}]^{i'}_{i}= \langle i'| {\cal V}|i\rangle$ and $\Phi_k^{mn}= E_k (t_m-t_n)$.
The factor $f^+_{\alpha k}$ is the number of particles in state $k$ of reservoir $\alpha$; 
it is $f^+_{\alpha k} =1 \big/\left(\e^{\delta S_{\alpha k}} + \nu\right)$ with $\nu=1$ for fermionic reservoirs and $\nu=-1$ for bosonic reservoirs.
Here,\cite{Footnote:units}
\begin{eqnarray}
\delta S_{\alpha k} = (E_k-\mu_\alpha)/T_\alpha \ ,
\label{Eq:deltaS_alphak}
\end{eqnarray}
which is the entropy change of reservoir $\alpha$ when a particle is added to state $k$.  
Identifying Eq.~(\ref{Eq:deltaS_alphak}) with an entropy change follows from the Claussius definition of entropy, applicable here because each reservoir is in its own local thermodynamics equilibrium with a well defined temperature.

The weight of $D_{a-}$ (for $a=0,1,2$) is given by
the Hermitian conjugate of $D_{a+}$
(so ${\cal V}^+\leftrightarrow {\cal V}^-$ and $\rmi\Phi_k^{mn} \to -\rmi\Phi_k^{mn}$)  with 
$f^+_{\alpha k}$ replaced by $f^-_{\alpha k}$.  
Hence
\begin{eqnarray}
\left[D^{\alpha k}_{0-}\right]^{i'_{m},i'_{n}}_{i_m,i_n}
\! &=& \!\! 
-\left[{\cal V}^+_{\alpha k}(t_m)\right]^{i'_{m}}_{i_m}
\left[{\cal V}^-_{\alpha k}(t_n)\right]^{i'_{n}}_{i_{n}} 
f^-_{\alpha k} \,\e^{\rmi \Phi_k^{nm}},
\label{Eq:D0-}
\\
\left[D^{\alpha k}_{1-}\right]^{j_m,i'_n}_{j'_m,i_n}
\! &=&\!  
\left[{\cal V}^+_{\alpha k}(t_m)\right]^{j_{m}}_{j'_{m}} 
\left[{\cal V}^-_{\alpha k}(t_n) \right]^{i'_{n}}_{ i_{n}} 
f^-_{\alpha k} \,\e^{ \rmi \Phi_k^{nm}} , \ 
\label{Eq:D1-}
\\
\left[D^{\alpha k}_{2-}\right]^{j_n,j_m}_{j'_n,j'_m}
\! &=& \!\!  
-\left[{\cal V}^+_{\alpha k}(t_n)\right]^{j_{n}}_{j'_{n}}
\left[{\cal V}^-_{\alpha k}(t_m)\right]^{j_{m}}_{j'_{m}}  
f^-_{\alpha k}\, \e^{\rmi \Phi_k^{nm}}. \qquad
\label{Eq:D2-}
\end{eqnarray}
Here $f^-_{\alpha k}$ is the number
of ways one can add a particle to state $k$ of reservoir $\alpha$.
For any reservoir (fermionic, bosonic or other) in internal equilibrium, 
\begin{eqnarray}
f^-_{\alpha k} =   \e^{\delta S_{\alpha k}} f^+_{\alpha k}\, ,
\label{Eq:local-detailled}
\end{eqnarray}
which is know as local detailed balance or micro-reversibility.
For fermion or boson distributions, this is guaranteed by the fact that $f^-_{\alpha k} = 1+ \nu f^+_{\alpha k}$ 
with $\nu=+1$ for fermions, and $\nu=-1$ for bosons.
Physically, $D^{\alpha k}_{1-}$ removes a particle from the system and adds it to state $k$ of reservoir 
$\alpha$; this adds a work of $\mu_\alpha$ and heat of $(E_k - \mu_\alpha)$ to reservoir $\alpha$. 
Thus, $D^{\alpha k}_{1-}$ involves a change of reservoir $\alpha$'s entropy of
$\delta S_{\alpha k}$ in Eq.~(\ref{Eq:deltaS_alphak}).
The reverse process, $D^{\alpha k}_{1+}$, removes such a particle from reservoir $\alpha$,
changing the reservoir's entropy by $-\delta S_{\alpha k}$. 
Contributions $D^{\alpha k}_{0\pm}$ and $D^{\alpha k}_{2\pm}$ 
do not change the number of particles in the reservoirs, and so involve no reservoir entropy change.

%%%%%%%%%%%%%%%%%%%%%%%%%%%%
\begin{figure}
\includegraphics[width=0.95\columnwidth]{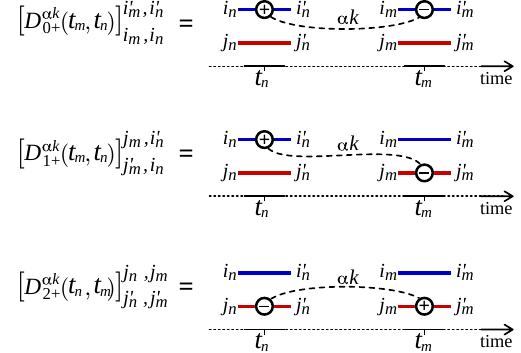}
\caption{\label{Fig:diagram-elements-2nd-order}
The second-order interaction with reservoir $\alpha$'s mode $k$.
Vertices marked ${\bm \oplus}$ or ${\bm \ominus}$ corresponds to the the matrices ${\cal V}^+_{\alpha k}$ or
${\cal V}^-_{\alpha k}$, respectively. 
The upper line is read from left to right, so the
${\bm \oplus}$ vertex in $D_{0+}$ or $D_{1+}$ indicates the matrix element
$[{\cal V}^+_{\alpha k}]_{i_{n+1}i_n}$. 
The lower line is read from right to left, so the ${\bm \oplus}$~vertex in $D_{2+}$ indicates
$[{\cal V}^+_{\alpha k}]_{j_{m}j_{m+1}}$.
Interaction $D_{a-}$ is given by $D_{a+}$ with ${\bm \oplus} \leftrightarrow {\bm \ominus}$, for $a=0,1,2$. 
}
\end{figure}
%%%%%%%%%%%%%%%%%%%%%%%%%%

\section{Total Entropy}   
The assumption of no Maxwell demons in the reservoirs implies that entanglement 
 between system and reservoir cannot be used to produce work. Then the correct definition of the total entropy production, $\Delta S_{\rm tot}$,  is the sum of that for the system (sys) and reservoirs (res), 
\begin{eqnarray}
\Delta S_{\rm tot} = \Delta S_{\rm sys} + \sum_\alpha \Delta S_{\rm res}^{(\alpha)}\, ,
\label{Eq:total_entropy}
\end{eqnarray}
with no term related to system-reservoir entanglement.
We take the change in entropy of each reservoir to be given by the Claussius formula.
This means that the change in reservoir $\alpha$'s entropy, $\Delta S_{\rm res}^{(\alpha)}$, for a trajectory $\dtraj$ 
is taken to be the sum of the entropy changes $\mp \delta S_{\alpha k}$ associated with each of the $D_{1\pm}^{\alpha k}$ transitions in $\dtraj$.   

As the system is typically in a highly non-equilibrium state, one cannot use the Claussius law to calculate its entropy.  
In the stochastic thermodynamics of classical systems,\cite{Schmiedl-Seifert2007,van-Broeck-review2015, our-review-sect8.10} an entropy is assigned to each system state
in such a way that the entropy of the system averaged over all such system states is the Shannon entropy.
For quantum systems, one can do exactly the same thing, if (and only if) the system's density matrix is in its diagonal basis.   
To get this entropy for the system's initial density matrix (at time $t_0$), we write it as
\begin{eqnarray}
\rho^{\rm sys}_{ml}(t_0)= \sum_{n_0} \big[{\cal W}_0\big]_{mn_0} \,P_{n_0}(t_0) \,\big[{\cal W}_0^\dagger\big]_{n_0l},
\end{eqnarray}
where ${\cal W}_0$ is the unitary matrix which rotates the system density matrix at time $t_0$ to its diagonal basis.
This means that $P_{n_0}(t_0)$ is the probability to find the system in state $n_0$ of its diagonal basis. 
In this basis, the system's von Neumann entropy, 
$-{\rm tr}\big[\rho_{\rm sys}(t_0) \ln [\rho_{\rm sys}(t_0)]\big]$ is simply  $-\sum_{n_0} P_{n_0}(t_0) \ln\big[P_{n_0}(t_0)\big]$,
where the sum is over the elements of the diagonal density matrix.
Thus, one can treat each element in the sum as a contribution to the entropy from a given initial state, so that state 
$n_0$'s contribution to the entropy of the initial system state is,
\begin{eqnarray}
S_{n_0}(t_0) &=& -\ln[P_{n_0}(t_0)],
\label{Eq:S_sys-ini}
\end{eqnarray}
with the average over all $n_0$ (i.e. a sum over $n_0$ weighted by the probability of state $n_0$) giving the system's initial von Neumann entropy.
The final system state's entropy (at time $t$) is calculated in the same way by rotating to the 
diagonal-basis of the final system density matrix, given by 
$\rho^{\rm sys}_{ml}(t)= \sum_n {\cal W}_{mn} \,p_n(t)\, \left[{\cal W}^\dagger\right]_{nl}$
for unitary ${\cal W}$, and assigning to the state $n$ an entropy of
\begin{eqnarray}
S_n(t) &=& - \ln[P_n(t)],
\label{Eq:S_sys-fin}
\end{eqnarray}
with $P_{n}(t)$ being the probability that the system is in state $n$ of the diagonal basis of its reduced density matrix at time $t$.
Eqs.~(\ref{Eq:S_sys-ini}.\ref{Eq:S_sys-fin}) can be used  
to associate trajectory $\dtrajdiag$, from initial state $n_0$ to final state $n$, with an entropy change in the system of   
\begin{eqnarray}
\Delta S_{\rm sys} (\dtraj) = S_n(t) - S_{n_0}(t_0) = -\ln\left[P_{n}(t) \over P_{n_0}(t_0)\right] ,
\label{Eq:DeltaS_sys}
\end{eqnarray}
as in the stochastic thermodynamics of classical rate equations.
Recall that this is only possible because the trajectory $\dtrajdiag$ is defined as going 
from a system-state $n_0$ in the diagonal basis
of the system density matrix at time $t_0$, to a system-state $n$ in the diagonal basis of the system's 
final density matrix (which is found by tracing out the reservoirs at the end of the evolution).  
This requires calculating the final density matrix (and finding its diagonal basis);
this is much like in usual  stochastic thermodynamics, 
where one also needs a complete knowledge of the final state probability distribution to assign entropies to it.

\section{First law of thermodynamics}
\label{Sect:firstlaw}

Here we show that energy conservation ensures that the first law of thermodynamics is obeyed on average.  If one goes beyond the average, there are fluctuations that violate the first law, much like the fluctuations that violate the second law.  These are little studied to-date, and merit a detailed study of their own. In this section, we restrict ourselves to considering the average energy in the set-up, and thereby show that the first law holds on average.

To ensure energy conservation, one must sum the three terms which contribute to the total energy; the energy in the reservoirs, the energy in the system, and the energy in the system-reservoir coupling.  If the system is not driven this total energy is conserved.  If the system is driven then the difference between the final and initial total energy is the work done by the drive, thus for between time $t_0$ and time $t$, the average work done by the drive is 
\begin{eqnarray}
\big\langle\Delta W_{\rm drive}(t;t_0)\big\rangle &=& \big\langle\Delta E_{\rm res}(t;t_0)\big\rangle  + \big\langle\Delta E_{\rm sys}(t;t_0)\big\rangle  
\nonumber \\
& & \ +\big\langle\Delta \Esr (t;t_0)\big\rangle  ,
\label{Eq:energy-conservation}
\end{eqnarray}
where $\Delta E_{\rm res}(t;t_0)$ is the energy change in the reservoirs,  
$\Delta E_{\rm sys}(t;t_0)= E_{\rm sys}(t)-E_{\rm sys}(t_0)$  is the energy change in the system,
and $\Delta \Esr (t;t_0) = \Esr (t)- \Esr (t_0)$ is the energy change in the system-reservoir coupling.

%%%%%%%%%%%%%%%%%%%%%%%%%%%%
\begin{figure}
\includegraphics[width=0.95\columnwidth]{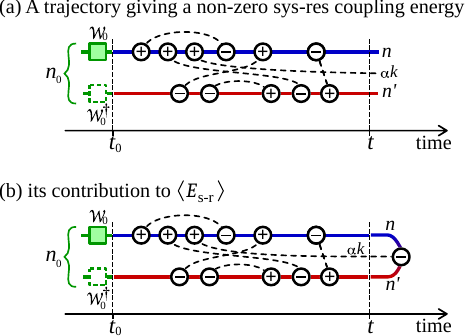}
\caption{\label{Fig:diagram-current}
(a) A trajectory which contributes to $[{\cal V}^-_{\alpha k}(t)]_n^{n'}\hat{c}^\dagger_{\alpha k}$, and (b) its contribution to the {\it average} energy in the system-reservoir coupling, $\langle \Esr\rangle$.
Such contributions to $\langle \Esr\rangle$  are the same as those 
necessary to calculate the current at time $t$, and their evaluation has been greatly discussed in 
Refs.~[\onlinecite{Schoeller-Schon1994,Konig96,Konig97,Schoeller-review1997}].
}
\end{figure}
%%%%%%%%%%%%%%%%%%%%%%%%%%

Of course, all terms in this sum are necessary to get to get energy conservation, irrespective of whether one can physically measure each of them or not.  Under the assumption of no Maxwell demons in the reservoirs,
made in section~\ref{Sect:no-demons}, one can measure the energy in the system and reservoirs, but not that in the system-environment coupling, since that depends on the state of individual reservoir modes.  In such a case, one could none the less determine $\big\langle\Delta\Esr(t;t_0)\big\rangle$ by using Eq.~(\ref{Eq:energy-conservation}), assuming one can also measure the work done by the drive, $\big\langle\Delta W_{\rm drive}(t;t_0)\big\rangle$.

The average energy in the quantum system is
\begin{eqnarray}
\big\langle E_{\rm sys}(t)\big\rangle &=& {\rm tr}_{\rm sys} \left[ \hat{H}_{\rm sys}(t) \rho_{\rm sys}(t)\right]  ,
\end{eqnarray}
while that in the system-reservoir coupling is
\begin{eqnarray}
\big\langle\Esr (t)\big\rangle&=& \sum_{\alpha \in {\rm el}} \! {\rm tr}\left[ \hat{V}_{\rm el}^{(\alpha)}(t)  \hat{\rho}_{\rm tot}(t)\right]
\nonumber \\
& &\qquad
+\sum_{\alpha \in {\rm ph}} \! {\rm tr}\left[ \hat{V}_{\rm ph}^{(\alpha)}(t)  \hat{\rho}_{\rm tot}(t)\right],
\label{Eq:E_sys}
\end{eqnarray}
where the trace in $\big\langle E_{\rm sys}\big\rangle$ is over the system states and $\rho_{\rm sys}(t)$ is the reduced system density matrix, but $\big\langle\Esr\big\rangle$ contains traces over the total density matrix (including reservoirs). 

The trajectories which sum to give the average change in the system energy, $\big\langle\Delta  E_{\rm sys}(t;t_0)\big\rangle$,
are those considered elsewhere in this article,
such as those in Fig.~\ref{Fig:sys+diagram}b 
or Fig.~\ref{Fig:rotate}.
However, the trajectories which sum to give the average change in the energy in the system-reservoir coupling, 
$\big\langle\Delta  \Esr(t;t_0)\big\rangle$, are rather different from those considered elsewhere in this article.
They have an additional single interaction vertex with mode  $k$ of reservoir $\alpha$ at some time before $t$, 
so that at time $t$ the system is in a superposition of a state with different numbers of particles in reservoir $\alpha$
(see Fig.~\ref{Fig:diagram-current}). 
Luckily, they have exactly the same structure as those used to calculate the current into the system in Refs.~\cite{Schoeller-Schon1994,Konig96,Konig97,Schoeller-review1997};
currents are given the difference between the term that creates a particle in the reservoir 
and one that destroys a particle, while the energy in the system-reservoir coupling is given by the sum of these two terms.
These trajectories are not discussed further here, because Ref.~[\onlinecite{Schoeller-Schon1994,Konig96,Konig97,Schoeller-review1997}] go into great detail about how to calculate their contribution.

The fact that reservoir $\alpha$ is in local thermodynamic equilibrium defined by its temperature $T_\alpha$ and electrochemical potential $\mu_\alpha$, mean it makes sense to split its energy, $\big\langle E_{\rm res}^{(\alpha)}(t)\big\rangle={\rm tr}\big[H_{\rm el/ph}^{(\alpha)}\hat{\rho}(t)\big]$, into two contributions
\begin{eqnarray}
\big\langle E_{\rm res}^{(\alpha)}(t)\big\rangle= -  \big\langle W_{\rm res}^{(\alpha)}(t)\big\rangle- \big\langle Q_{\rm res}^{(\alpha)}(t)\big\rangle,
\end{eqnarray}
with
\begin{eqnarray}
\big\langle W_{\rm res}^{(\alpha)}(t) \big\rangle &=& - {\rm tr}\left[ 
\mu_\alpha \sum_k \hat{c}_{\alpha k}^\dagger \hat{c}_{\alpha k}\ 
\hat{\rho}(t)\right], 
\\
\big\langle Q_{\rm res}^{(\alpha)}(t) \big\rangle &=&-{\rm tr}\left[ 
\sum_k \,\left(E_{\alpha k}-\mu_\alpha \right)\, \hat{c}_{\alpha k}^\dagger \hat{c}_{\alpha k} \ 
\hat{\rho}(t)\right]. \quad 
\end{eqnarray}
Then the change in reservoir energy between time $t_0$ and $t$ can be defined as
\begin{eqnarray}
\big\langle \Delta E_{\rm res}^{(\alpha)}(t;t_0)\big\rangle 
= -\big\langle \Delta W_{\rm res}^{(\alpha)}(t;t_0)\big\rangle
-\big\langle \Delta Q_{\rm res}^{(\alpha)}(t;t_0)\big\rangle,
\end{eqnarray}
the first quantity here is the average work done by the reservoir 
$\big\langle \Delta W_{\rm res}^{(\alpha)}(t;t_0)\big\rangle 
= \big\langle W_{\rm res}^{(\alpha)}(t)\big\rangle  - \big\langle W_{\rm res}^{(\alpha)}(t_0)\big\rangle$,
and the second quantity is the average heat flow out of the reservoir
$\big\langle \Delta Q_{\rm res}^{(\alpha)}(t;t_0)\big\rangle 
= \big\langle Q_{\rm res}^{(\alpha)}(t)\big\rangle  - \big\langle Q_{\rm res}^{(\alpha)}(t_0)\big\rangle$.
There is no ambiguity in this separation, because the fact the reservoir is in local equilibrium means that 
the former (the work done) has no entropy change associated with it, while the latter (the heat change) is associated with an entropy change of 
\begin{eqnarray}
\Delta S_{\rm res}^{(\alpha)} =  \Delta Q_{\rm res}^{(\alpha)}\big/T_\alpha.
\label{Eq:res-S&Q}
\end{eqnarray}
The trajectories which sum to give these average changes in a reservoir's energy are those considered elsewhere in this article,
such as those shown in Fig.~\ref{Fig:sys+diagram}b 
or Fig.~\ref{Fig:rotate}.
However the change in work or heat in the reservoir $\alpha$ is extremely easy to read from a given trajectory,
one simply sums up the change in work or heat for each dashed line symbolizing $D_{1\pm}^{\alpha k}$, 
as outlined at the end of section~\ref{Sect:trajectories}.

Given these definitions and Eq.~(\ref{Eq:energy-conservation}), one easily arrives at the first law of thermodynamics
for the average dynamics of the set-up;
\begin{eqnarray}
& & \hskip -10mm \big\langle\Delta E_{\rm sys}(t;t_0)\big\rangle  
+ \big\langle\Delta \Esr(t;t_0)\big\rangle  
\nonumber \\
&=& \big\langle\Delta W(t;t_0)\big\rangle  +\sum_\alpha \big\langle \Delta Q_{\rm res}^{(\alpha)} (t;t_0)\big\rangle , \quad\quad
\label{Eq:1stlaw}
\end{eqnarray}
where we define $ \Delta W(t;t_0)$ as the total work done on the system by drive or reservoirs,
\begin{eqnarray}
\big\langle\Delta W(t;t_0)\big\rangle = \big\langle\Delta W_{\rm drive}(t;t_0)\big\rangle   + \sum_\alpha \big\langle \Delta W_{\rm res}^{(\alpha)}(t;t_0)\big\rangle . \quad
\end{eqnarray}
It thereby seems natural to interprete $\big\langle\Delta E_{\rm sys}(t;t_0)\big\rangle  + \big\langle\Delta \Esr(t;t_0)\big\rangle $ as the change in the {\it effective} internal energy of the system (an effective energy which includes the system-reservoir coupling), as mentioned in section~\ref{Sect:comment-sys-res}.

Just as in classical thermodynamics systems, the simplest cases to consider are those where the system returns to its initial state at the end of the evolution, 
so that its internal energy is the same at the final time, $t$, as it was at the initial time, $t_0$. 
Then $\big\langle\Delta E_{\rm sys}(t;t_0)\big\rangle=\big\langle\Delta \Esr(t;t_0)\big\rangle=0$, which means that 
Eq.~(\ref{Eq:1stlaw}) directly gives the simplest and best-known consequence of the first law; {\it the work output of the machine equals the heat absorbed from the reservoirs}.

Note that the change of energy in the reservoirs was separated into a change of heat and a change of work,
but this was not done for the energy of the system or the system-reservoir coupling. The reason is that each reservoir is in local thermodynamic equilibrium, with a well-defined temperature,
when the system and system-reservoir couplings are typically far from equilibrium with no well-defined temperature.
Thus there is no ambiguity in the separation of energy into heat and work in a reservoir, see Eq.~(\ref{Eq:res-S&Q}),
but there is no simple way to make the same separation for the system or for the system-reservoir couplings.

\section{Time-reversed set-up}
\label{Sect:TR-derivation}

As in classical systems, one derives fluctuation theorems by comparing two different set-ups (A and B), where the Hamiltonian in set-up B  is the time-reverse of the Hamiltonian in set-up A over the time-window from $t_0$ to $t$.  
To be clear, whatever the Hamiltonian of set-up A, we can invent a set-up B whose Hamiltonian is the  time-reverse of set-up A.
In the special case of a time-independent Hamiltonian without external magnetic fields or spins,
the two set-ups are identical, but otherwise they are not.

The objective of this section is to make the connection between weight of trajectories on the Keldysh contour in 
set-up B and set-up A.  This starts by making the connection between the terms in the Hamiltonians of set-ups A and B in section~\ref{Sect:H-TR}, and then between the perturbative terms in the interaction representation in 
section~\ref{Sect:interaction-TR}.  This enables one to make the connection between the weight of individual transitions in section~\ref{Sect:D-TR}, from which one gets the  connection between weight of trajectories on the Keldysh contour in set-up B and set-up A in section~\ref{Sect:Traj-TR}.  The {\it central observation} of this work is this relationship, given in Eq.~(\ref{Eq:central-obs}) below. 
It is this relationship which is so similar to the relationship between trajectories in the stochastic thermodynamics theory
 for 
classical Markovian systems,\cite{Schmiedl-Seifert2007,van-Broeck-review2015, our-review-sect8.10} that we can use very similar logic to derive various well-known fluctuation theorems in section~\ref{Sect:fluct}.

\subsection{Time-reversed Hamiltonian}
\label{Sect:H-TR}

%%%%%%%%%%%%%%%%%%%%%%%%%%%%
\begin{figure}
\includegraphics[width=\columnwidth]{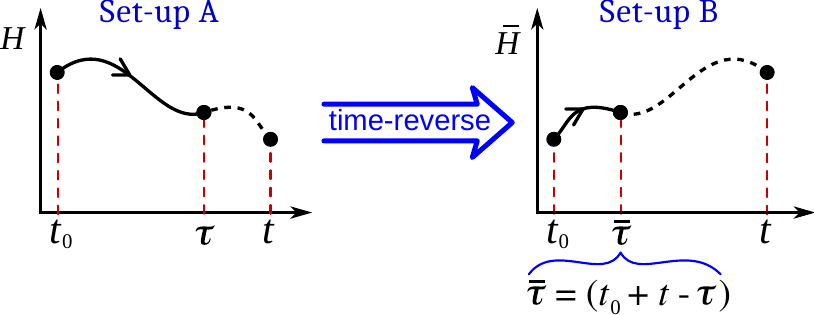}
\caption{\label{Fig:a1}
A sketch of how time-reversal affects the Hamiltonian, in the absence of external magnetic fields and spins.
 If there are external magnetic fields and spins, then the time-reverse is given in Appendix~\ref{Append:TR}. }
\end{figure}
%%%%%%%%%%%%%%%%%%%%%%%%%%

If A's Hamiltonian (system+reservoir) is $\hat{H}$ in Eq.~(\ref{Eq:H_tot}), then B's is $\overline{\hat{H}(\tau)} = \hat{\Theta}^\dagger \hat{H}(t_0+t-\tau) \hat{\Theta}$, where $\hat{\Theta}$ is the time-reverse operator in Messiah's texbook.\cite{Messiah-chapt15}   The main results we need from Messiah's texbook are recalled here in Appendix~\ref{Append:TR}, with the most trivial case sketched in Fig.~\ref{Fig:a1}. 
Thus if set-up A has a given time-dependent system Hamiltonian, $H_{\rm sys}(\tau)$, with given system-reservoir couplings, $V^\pm_{\alpha k}(\tau)$, then set-up B is {\it chosen}  to have 
the system Hamiltonian and system-reservoir couplings
\begin{eqnarray}
\overline{\hat{H}_{\rm sys}(t_0+t-\tau)} &=& \hat{\Theta}^\dagger \hat{H}_{\rm sys}(\tau) \hat{\Theta}
\\
\overline{\hat{V}^\pm_{\alpha k}(t_0+t-\tau)} &=&  \hat{\Theta}^\dagger \hat{V}^\pm_{\alpha k}(\tau) \hat{\Theta}
\label{Eq:overbar-V}
\end{eqnarray}
where the bar above a symbol means that it is in set-up B, 
while the bar's absence means in it is in set-up A.

These equations are cast in terms of matrix elements by inserting them between $\langle \overline{\imath}|=\langle i|\hat{\Theta}$ and $|\overline{\jmath}\rangle =\hat{\Theta}^\dagger|j\rangle$, then
\begin{eqnarray}
\left[\ \overline{H_{\rm sys}(t_0+t-\tau)}\right]^{\overline{\imath}}_{\overline{\jmath}} 
&=&\left[ H_{\rm sys}(\tau) \right]^i_j
\\
\left[\ \overline{V^\pm_{\alpha k}(t_0+t-\tau)}\right]^{\overline{\imath}}_{\overline{\jmath}} 
&=& \left[ V^\pm_{\alpha k}(\tau) \right]^i_j
\end{eqnarray}
where $
\left[\cdots\right]^{\overline{\imath}}_{\overline{\jmath}} =\langle \overline{\imath}| \cdots | \overline{\jmath} \rangle$.
Thus the matrix elements for transitions from system state $|\overline{\jmath}\rangle$
to system state $|\overline{\imath}\rangle$ in set-up B (whose Hamiltonian is the time-reverse of set-up A's), 
are the same as the 
matrix elements from system state $|j\rangle$ to system state $|i\rangle$ in set-up A.
 Spinless systems written in a basis of position states are trivial, because then $ | \overline{\imath} \rangle = | i \rangle$. However, if one is working with basis states with non-zero momentum states, then 
 $ | \overline{\imath} \rangle$ is the state with the opposite momentum from $| i \rangle$.
 If one is working with spins, then the state $ | \overline{\imath} \rangle$ is the state with the opposite spin from from state $| i \rangle$.
 
Equally, the reservoir Hamiltonians for set-up B are the time-reverse of those in set-up A,
so reservoir $\alpha$ in set up B has a Hamiltonian
\begin{eqnarray}
\overline{\hat{H}_{\rm el}^{(\alpha)}} = \hat{\Theta}^\dagger \hat{H}_{\rm el}^{(\alpha)} \hat{\Theta}
\label{Eq:overline-H_el}
\end{eqnarray}
where $\hat{H}_{\rm el}^{(\alpha)}$ is that reservoir's Hamiltonian in set-up A.
In the absence of spins or external magnetic fields, this time-reverse operation is of no consequence. 
However, if reservoir $\alpha$ in set-up A is a reservoir of electrons which are spin-up with respect to some axis, then the same reservoir in set-up B will contain electrons which are spin-down with respect to that axis.
Similarly, if there is an external magnetic field acting on the reservoir in set-up A, then the field must be reversed
in that reservoir in set-up B.
For photon or phonon reservoirs, the relation between their Hamiltonians in the two set-ups is the same 
as in Eq.~(\ref{Eq:overline-H_el}).

\subsection{Reservoir states are not time-reversed}
\label{Sect:non-reversal_of_res_states}

If one evolves an initial state under a Hamiltonian,
time-reverses the state,  and evolves it under the time-reversed Hamiltonian, 
the dynamics in the second part of the evolution will look like a time reverse of the dynamics in the first part of the evolution.

However, this work's set-up A and set-up B are a different time-reversal situation, 
in which each set-up is divided into a system and reservoirs,
and we make the ``assumption of no Maxwell demons in the reservoirs'' in section~\ref{Sect:no-demons}.
This assumes the set-up and its drive are not aware of the microscopic dynamics of the reservoirs,
as such  one {\it cannot  time-reverse the state of the individual modes in the reservoirs}, even if one can 
can time-reverse the reservoirs' Hamiltonians (typically time-reversing the reservoir part of the Hamiltonian only requires interchanging the chemical potentials on spin-up and spin-down reservoirs and reversing any external magnetic fields acting on the reservoirs).
Thus, even if we time-reverses the system state and time-reverses the total Hamiltonian, we will not see time-reversed dynamics, because we have not time-reversed the reservoir states. 

Suppose set-up A starts with the system and reservoirs in a produce state, and then evolves. 
The system becomes correlated and/or entangled with individual reservoir modes.
A measurement of  the system state indicates that it is decohering and decaying towards a thermal state.
A measurement of individual reservoir modes shows that an infinitesimal proportion of them are acquiring a non-thermal state.
Then in set-up B, the total Hamiltonian is the time-reverse of that is set-up A, 
and the initial state is a product state, where the system state is the time reverse of the final system state in set-up A, and the
reservoir modes are taken to be thermal (i.e.\ not time-reversed).
The system state in set-up B does {\it not} become less correlated and/or entangled with the reservoirs as it evolves (as it would if we had time-reversed the full state, including the reservoir modes). Instead, the system continues to become more entangled with reservoir modes, which means that  a measurement of the system state in set-up B will indicate that it also decoheres and decays towards a thermal state.

\subsection{Time-reversal for the interaction representation}
\label{Sect:interaction-TR}

Under time-reversal, the matrix representation of the system evolution operator is
\begin{eqnarray}
\overline{{\cal U}_{\rm sys}(t+t_0-\tau;t_0)} = \hat{\Theta}^\dagger \, {\cal U}_{\rm sys}^\dagger(t;\tau)  \, \hat{\Theta}.
\end{eqnarray}
Now, to simplify the algebra, it is assumed that a complete solution of the dynamics under $H_{\rm sys}$  exists, 
then the final state of the system (its state at given time $t$) can always be written in a basis chosen such that ${\cal U}_{\rm sys}(t;t_0) = 1$. 
Then, the unitary of ${\cal U}_{\rm sys}$ means that
\begin{eqnarray}
{\cal U}_{\rm sys}^\dagger (t;\tau)\ =\ {\cal U}_{\rm sys}(\tau;t_0)  \, .
\label{Eq:U_is_1}
\end{eqnarray}
Given this one has
\begin{eqnarray}
\overline{{\cal U}_{\rm sys}(t+t_0-\tau;t_0)} = \hat{\Theta}^\dagger \ {\cal U}_{\rm sys}(\tau;t_0) \ \hat{\Theta}\, .
\label{Eq:using_fact_that_U_is_1}
\end{eqnarray}
The interaction between the system and the reservoirs at time $\overline{\tau}=(t_0+t-\tau)$
is written in the interaction representation for the {\it time-reversed Hamiltonian} as 
\begin{eqnarray}
 \overline{{\cal V}^\pm_{\alpha k} (\overline{\tau})}
&=& \overline{{\cal U}_{\rm sys}^\dagger(\overline{\tau};t_0)} \ \,\overline{V^\pm_{\alpha k} (\overline{\tau})} \ \,\overline{{\cal U}_{\rm sys}(\overline{\tau};t_0)} 
\end{eqnarray}
where the matrix $\overline{V^\pm_{\alpha k} (\overline{\tau})}$ is defined above.
Substituting in Eqs.~(\ref{Eq:overbar-V},\ref{Eq:using_fact_that_U_is_1}) on the right, 
and comparing with Eq.~(\ref{Eq:V-interaction-representation}), one finds that
\begin{eqnarray}
 \overline{{\cal V}^\pm_{\alpha k} (t_0+t-\tau)}
&=& \hat{\Theta}^\dagger \ {\cal V}^\pm_{\alpha k} (\tau) \ \hat{\Theta} 
\end{eqnarray}
for all $\tau$ between $t_0$ and $t$.

For what follows it is convenient to cast this equality in terms of matrix elements by inserting it between $\langle \overline{\imath}|=\langle i|\hat{\Theta}$ and $|\overline{\jmath}\rangle =\hat{\Theta}^\dagger|j\rangle$, then
\begin{eqnarray}
\left[\,\overline{{\cal V}^\pm_{\alpha k} (t_0+t-\tau)}\,\right]^{\overline{\imath}}_{\overline{\jmath}} 
\ =\ \left[\,{\cal V}^\pm_{\alpha k} (\tau)\,\right]^i_j 
\label{Eq:V-TR-matrix-elements}
\end{eqnarray}
Thus the matrix element for reservoir-induced transitions from system state $|\overline{\jmath}\rangle$
to system state $|\overline{\imath}\rangle$ in set-up B (the set-up whose Hamiltonian is the time-reverse of set-up A's), is the same as the 
matrix element from system state $|j\rangle$ to system state $|i\rangle$ in set-up A.

%====================
\subsection{Time-reversal symmetry between $D_{a\pm}^{\alpha k}$-transitions}
\label{Sect:D-TR}

Eq.~(\ref{Eq:D1+}) implies that the $D_{1+}$ transition in the time-reserved system 
(set-up B) must have the weight
\begin{eqnarray}
\left[\overline{D^{\alpha k}_{1+}(\overline{t}_m,\overline{t}_n)}\right]^{\overline{\jmath}_m,\overline{\imath}'_n}_{\overline{\jmath}'_m,\overline{\imath}_n}
\! &=&\! 
\left[\overline{{\cal V}^-_{\alpha k}(\overline{t}_m)} \right]^{\overline{\jmath}_m}_{\overline{\jmath}'_m}
\left[\overline{{\cal V}^+_{\alpha k}(\overline{t}_n)} \right]^{\overline{\imath}'_n}_{\overline{\imath}_n} \nonumber  \\
& & \times
f^+_{\alpha k} \,\exp\left[ \rmi E_k (\overline{t}_m-\overline{t}_n) \right], \qquad 
\end{eqnarray}
where $\overline{t}_n= t_0+t-t_n$.
Now substituting in Eq.~(\ref{Eq:V-TR-matrix-elements}), and noting that 
$(\overline{t}_m-\overline{t}_n) = (t_n-t_m)$, we get
\begin{eqnarray}
\left[\overline{D^{\alpha k}_{1+}(\overline{t}_m,\overline{t}_n)}\right]^{\overline{\jmath}_m,\overline{\imath}'_n}_{\overline{\jmath}'_m,\overline{\imath}_n}
\! &=&\! 
\left[{\cal V}^-_{\alpha k}(t_m) \right]^{j_m}_{j'_m}
\left[{\cal V}^+_{\alpha k}(t_n) \right]^{i'_n}_{i_n} \nonumber  \\
& & \times
f^+_{\alpha k} \,\exp\left[ \rmi E_k (t_n-t_m) \right], \qquad 
\end{eqnarray}
Now comparing this with $D_{1-}$ in Eq.~(\ref{Eq:D1-}), one sees the only difference is the factors 
of $f^\pm_{\alpha k}$.  However, local detailed balance in reservoir $\alpha$ implies Eq.~(\ref{Eq:local-detailled}), so
\begin{eqnarray}
\left[\overline{D^{\alpha k}_{1+}(\overline{t}_m,\overline{t}_n)}\right]^{\overline{\jmath}_m,\overline{\imath}'_n}_{\overline{\jmath}'_m,\overline{\imath}_n}
\! &=&\! 
\left[D^{\alpha k}_{1-}(t_m, t_n)\right]^{j_m,i'_n}_{j'_m,i_n} \e^{-\delta S_{\alpha k}}. \qquad 
\end{eqnarray}
Exactly the same logic holds if one starts with $\overline{D^{\alpha k}_{1-}}$ in place of $\overline{D^{\alpha k}_{1+}}$.  One just has to take the Hermitian conjugate throughout 
(so ${\cal V}^+\leftrightarrow {\cal V}^-$ and $\rmi\Phi_k^{mn} \to -\rmi\Phi_k^{mn}$)  and replace 
$f^+_{\alpha k}$ by $f^-_{\alpha k}$, getting the results in Eq.~(\ref{Eq:D_1-TR}).

Similarly, Eq.~(\ref{Eq:D0+}) means that 
\begin{eqnarray}
\left[\overline{D_{0+}^{\alpha k}(\overline{t}_n,\overline{t}_m)}\right]
^{\overline{\jmath}_n, \overline{\jmath}_m}_{\overline{\jmath}'_n,\overline{\jmath}'_m}
&=& -\left[\overline{{\cal V}^-_{\alpha k}(\overline{t}_n)}\right]^{\overline{\jmath}_n}_{\overline{\jmath}'_n}
 \left[\overline{{\cal V}^+_{\alpha k}(\overline{t}_m)}\right]^{ \overline{\jmath}_m}_{\overline{\jmath}'_m} 
\nonumber \\
& & \times \,
 f^+_{\alpha k} \,\exp\left[{\rmi E_k(\overline{t}_n-\overline{t}_m)}\right], \qquad
\end{eqnarray}
note that $\overline{t}_n > \overline{t}_m$, since it is assumed that $t_m>t_n$.
As above, Eq.~(\ref{Eq:V-TR-matrix-elements}) is substituted in, and one notes that 
$(\overline{t}_m-\overline{t}_n) = (t_n-t_m)$, to get
\begin{eqnarray}
\left[\overline{D_{0+}^{\alpha k}(\overline{t}_n,\overline{t}_m)}\right]
^{\overline{\jmath}_n, \overline{\jmath}_m}_{\overline{\jmath}'_n,\overline{\jmath}'_m}
&=& -\left[{\cal V}^-_{\alpha k}(t_n)\right]^{j_n}_{j'_n}
 \left[{\cal V}^+_{\alpha k}(t_m)\right]^{j_m}_{j'_m} 
\nonumber \\
& & \times \,
 f^+_{\alpha k} \,\exp\left[{\rmi E_k(t_m-t_n)}\right], \qquad
\end{eqnarray}
which is the same as the right hand side of Eq.~(\ref{Eq:D2+}).
One can do the same for $\overline{D_{0-}^{\alpha k}}$.

The result of all these relations between $D$s in the time-reversed set-up (set-up B) and the original set-up (set-up A)
can be summarized as follows
\begin{subequations}
\label{Eq:D_0,1,2-TR}
\begin{eqnarray}
\left[\overline{D_{1\mp}^{\overline{\alpha} k}(\overline{t}_m,\overline{t}_n)}\right]^{\overline{\jmath}_m,\overline{\imath}'_n}_{\overline{\jmath}'_m,\overline{\imath}_n}
 &=&\! \left[D_{1\pm}^{\alpha k}(t_n,t_m)\right]^{i'_n,j_m}_{i_n,j'_m} \ \e^{\pm \delta S_{\alpha k}}\, , \qquad
\label{Eq:D_1-TR}
 \\
\left[\overline{D_{0\pm}^{\overline{\alpha} k}(\overline{t}_n,\overline{t}_m)}\right]
^{\overline{\jmath}_n, \overline{\jmath}_m}_{\overline{\jmath}'_n,\overline{\jmath}'_m}
&=& \! \left[D_{2\pm}^{\alpha k}(t_n,t_m)\right]^{j_n,j_m}_{j'_n,j'_m} \ \ ,
\label{Eq:D_0,2-TR}
\end{eqnarray}
\end{subequations}
where $\overline{t}_n= t_0+t-t_n$.  

\subsection{Time-reversed trajectories}
\label{Sect:Traj-TR}

\begin{figure}
\includegraphics[width=0.95\columnwidth]{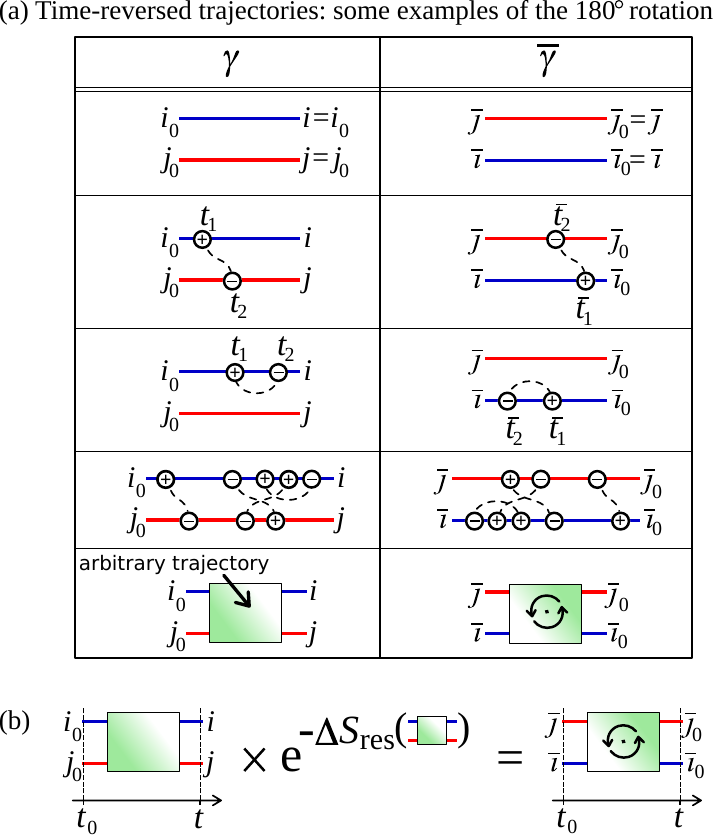}
\caption{\label{Fig:rotate}
(a) Time-reversal of trajectories on the Keldysh contour, via a $180^\circ$ rotation 
in the plane of the page.  The interaction times are related by $\overline{t}_n=t+t_0-t_n$.
(b) A graphical representation of Eq.~(\ref{Eq:central-obs}), where the shaded box is the trajectory's weight, $P(\dtraj)$, and its $180^\circ$ rotation is $\overline{P}(\overline{\dtraj})$.
}
\end{figure}

For any trajectory, $\dtraj$, on the Keldysh contour in set-up A, 
one can define a trajectory  $\overline\dtraj$ in set-up B which is the time-reverse of $\dtraj$.
More precisely, $\overline\dtraj$ is defined by rotating $\dtraj$ by $180^\circ$ in the plane of the page, and replacing all states by their time reverse, see Fig.~\ref{Fig:rotate}a.
The time-reverse of state $|i_n\rangle$  is $|\overline{\imath}_n\rangle = \hat{\Theta}^\dagger |i_n\rangle$.
% so $|\overline{\imath}_n\rangle$ is the same as $ |i_n\rangle$ with momentum and spin reversed.

One then observes that  if $\dtraj$ 
contains a $D$-factor on the right hand side of one of the equality in Eq.~(\ref{Eq:D_0,1,2-TR}), then 
then $\overline{\dtraj}$ contains the $D$-factor on the left hand side of the same equality, and vice-versa.
The weights of trajectory $\dtraj$ in set-up A and $\overline{\dtraj}$ in set-up B
are given by products 
of the factors of $D_{a\pm}^{\alpha k}$ that form each of them, this results in 
the {\it central observation} of this work (shown graphically in  Fig.~\ref{Fig:rotate}b),
\begin{eqnarray}
\overline{P} \big(\overline{\dtraj}\big) = P\big(\dtraj\big) 
\exp\big[-\Delta S_{\rm res}\big(\dtraj\big) \big],
\label{Eq:central-obs}
\end{eqnarray}
where $P(\dtraj)$ is the weight of double-trajectory $\gamma$ in set-up A,  
and $\overline{P}(\overline{\dtraj})$ is that of $\overline{\gamma}$ in set-up B.
The  reservoir entropy change, $\Delta S_{\rm res} (\dtraj)$, is the sum of the $\delta S_{\alpha k}$ for all  transitions in $\dtraj$.

Now consider a double-trajectory, $\dtrajdiag$, 
which goes from the $n_0$th state in the diagonal basis of $\rho(t_0)$
to the $n$th state in the diagonal basis of $\rho^{\rm sys}_{ml}(t)$ (see Fig.~\ref{Fig:sys+diagram}b).
The subscript ``d'' is to indicate that it goes from diagonal basis to
diagonal basis.  Let us define its weight as $P(\dtrajdiag)$, this equals $P(\dtraj)$ multiplied by a factor of  $\big[{\cal W}_0\big]_{i_0n_0} [{\cal W}_0^\dagger]_{n_0j_0}$ to transformation out of the diagonal basis at time $t_0$, and a factor of 
$[{\cal W}^\dagger]_{ni}{\cal W}_{jn}$ to go to the diagonal basis at time $t$.
The unitarity of the transformations ${\cal W}_0$ and ${\cal W}$ means they do not change $\Delta S_{\rm res}$ or $\Delta S_{\rm sys}$, so one has
\begin{eqnarray}
\overline{P} \big(\overline{\dtrajdiag}\big) = P\big(\dtrajdiag\big) \exp[-\Delta S_{\rm res}(\dtrajdiag)].
\label{Eq:central-obs-diag}
\end{eqnarray}
Here, one must recall that the trajectory $\dtrajdiag$ is in set-up A, and goes
from the state $n_0$ in the diagonal basis of the (initial) system density matrix at time $t_0$,
to the state $n$ in the diagonal basis of the (final) system's reduced density matrix at time $t$.
Its time-reverse trajectory $\overline{\dtrajdiag}$ is a trajectory in set-up B which goes from 
state $\overline{n}$ at time $t_0$ to state $\overline{n_0}$ at time $t$.

This relation is much the same as in Markovian stochastic thermodynamics of classical systems
and which was used to derive  various of the best known fluctuation theorems.\cite{Schmiedl-Seifert2007,van-Broeck-review2015, our-review-sect8.10}
In the next section, we will show that a very similar procedure allows us to derive these fluctuation theorems for 
quantum systems with non-Markovian dynamics.
In particular, we will show that the integral fluctuation theorem in Eq.~(\ref{Eq:non-equil_partition_identity})  holds for any system, 
which we will show implies that any system will satisfy the second law of thermodynamics on average, $\big\langle \Delta S_{\rm tot} \big\rangle \geq 0$.

However, a complication in these quantum system (absent in the classical ones)  is the question of the basis in which the dynamics are diagonal. 
It is crucial to note that the bases in which we consider the states $\overline{n}$ and $\overline{n_0}$
in set-up B, are those defined as the basis in which set-up A's final and initial system density matrices are diagonal.   In general, these will not both coincide with the bases in which the system density matrix for set-up B will be diagonal.  Consider the initial state in set-up B which coincides with the time-reverse of the final state in set-up A; it  will not evolve to a state that coincides with the initial state in set-up A (cf.\ section~\ref{Sect:non-reversal_of_res_states}).  Thus, there is no reason to expect the final state in set-up B to be diagonal in the same basis as the initial state in set-up A.
In this case $\overline{n_0}$ corresponds to the $n_0$th diagonal matrix element in the reduced system density matrix in set-up B, when that density matrix is {\it not} written in its diagonal basis, 
but is written in the basis in which set-up A's initial density matrix was diagonal.

This is not a problem in deriving certain fluctuation theorems, such as Eq.~(\ref{Eq:non-equil_partition_identity}).
However, our derivation of the Crooks equation only apply in those special cases
in which the final state in set-up B is \dRed{\sout{diagonal {\it in the same basis}} the same} as the initial state in set-up A; section~\ref{Sect:Crooks} elaborates on this point, and gives examples of special cases for which it applies.

%%%%%%%%%%%%%%%%%%%%%%%%%%%%
%%%%%%%%%%%%%%%%%%%%%%%%%%

\section{Fluctuation theorems}
\label{Sect:fluct}

Schmiedl and Seifert showed in Ref.~[\onlinecite{Schmiedl-Seifert2007}] that trajectories in classical rate equations obey 
Eq.~(\ref{Eq:central-obs}), and one can derive most of the standard fluctuation relations from 
suitable sums over these classical trajectories.
Their proofs were for a discrete set of states with transitions governed by Markovian rate equations.
Together with a more complicated continuum version,\cite{Seifert-PRL2005} this became known as {\it stochastic thermodynamics}, and it is discussed in a number of reviews.\cite{Seifert-review2012,van-Broeck-review2015, our-review-sect8.10}  
Our objective here is to show that the same logic as in Refs~[\onlinecite{Schmiedl-Seifert2007,van-Broeck-review2015, our-review-sect8.10}] can be used to derive fluctuation theorems from the Keldysh-contour trajectories using Eq.~(\ref{Eq:central-obs-diag}).
Before going into the detail of the derivations for non-Markovian quantum systems, 
which are very close to the derivations for classical rate equations in Refs.~[\onlinecite{Schmiedl-Seifert2007,van-Broeck-review2015, our-review-sect8.10}], 
we mention the points which differ between stochastic thermodynamics for classical rate equations
and for non-Markovian quantum systems.

The most obvious difference is that the trajectories themselves are very different.
The quantum system's trajectories come from perturbation theory on the Keldysh contour, 
while the trajectories in Refs.~[\onlinecite{Schmiedl-Seifert2007,van-Broeck-review2015, our-review-sect8.10}] come from classical rate processes.  
One consequence of this is a trajectory  $\dtrajdiag$, on the Keldysh contour, typically has 
a complex weight $P(\dtrajdiag)$.
However, every trajectory has a partner with the same entropy change, but with the complex conjugate weight; this trajectory is found from its partner by interchanging the trajectory's upper and lower lines and taking ${\bm \oplus} \leftrightarrow {\bm \ominus}$.
Any physical probability will involve an equal sum of the two weights, and so will be real.
None the less, this sum of a trajectory and it complex conjugate partner will often be a negative real number,
so it should be considered as a contribution to the probability, and not a probability itself.
The contributions with negative weights reduce the probability to go to a given state, 
while those with positive weights increase the probability to go to another state.

These negative weights do not occur in the usual stochastic thermodynamics of classical rate equations,
however it is easy to see why.
In usual stochastic thermodynamics, the probability that a trajectory in state $i$ has no transitions 
in the time window $\tau_n$ to $\tau_{n+1}$ is  \cite{Schmiedl-Seifert2007,van-Broeck-review2015, our-review-sect8.10}
$\exp\big[-\int_{\tau_n}^{\tau_{n+1}} \rmd \tau \,\Gamma_{i}(\tau) \big]$, 
where $\Gamma_{i}(\tau)$ is the sum of all transition rates out of state $i$ at time $\tau$.  
To compare this with our quantum theory (which is perturbative in the reservoir couplings), such
exponential terms should be expanded in powers of $\Gamma_i$.  This generates a version of stochastic thermodynamics in which trajectories can have positive or negative weights. 
Our quantum theory has trajectories with positive and negative weights for the same reason.

The weight of a trajectory $\dtrajdiag$ obeys Eq.~(\ref{Eq:central-obs-diag}), 
combining this with Eq.~(\ref{Eq:DeltaS_sys}) gives
\begin{eqnarray}
\overline{P} \big(\overline{\dtrajdiag}\big) \,P_{n}(t) &=& P\big(\dtrajdiag\big)\, P_{n_0}(t_0) \,\exp[-\Delta S_{\rm tot}(\dtrajdiag)] ,\qquad
\label{Eq:fluct-rel-ingredient}
\end{eqnarray}
where $\Delta S_{\rm tot}(\dtrajdiag)$ is the sum of the entropy change in system and reservoirs
(see Eq.~(\ref{Eq:total_entropy})) associated with trajectory $\dtrajdiag$ from state $n_0$ at time $t_0$ to state $n$ at time $t$.
Despite the difference in the nature of the trajectories, 
this relation is the same as for classical rate equations, where t was used to 
derive various well-known fluctuation theorems.  Now, we can follow basically the same derivations 
to derive the same fluctuation relations for non-Markovian quantum systems.
These derivations are presented in the following subsections, 
readers familiar with Refs.~[\onlinecite{Schmiedl-Seifert2007,van-Broeck-review2015, our-review-sect8.10}] 
will notice that their similarity to those for classical rate equations.

\subsection{Integral fluctuation theorem}
\label{Sect:NEPI}

Let us start by deriving Eq.~(\ref{Eq:non-equil_partition_identity}), which 
is known as the {\it non-equilibrium partition identity} \cite{Yamada-Kawasaki1967,Morriss1985,Carberry2005} 
as well as the {\it integral fluctuation theorem}.\cite{Schmiedl-Seifert2007,van-Broeck-review2015}
In classical systems it is the most general fluctuation theorem, since one can use stochastic thermodynamics
to show that it applies to any classical system with Markovian dynamics, irrespective of that system's initial or final state.
This section will show that the same is true for non-Markovian quantum systems.

If one has a physical quantity (energy, particle current, entropy or similar)
that one can calculate for each trajectory of the system, then the average value of that quantity for a system is 
given by the following sum over all trajectories 
\begin{eqnarray}
\left\langle \cdots \right\rangle &=&  
\sum_{n_0,n} \ \sum_{\dtrajdiag \in  \{n_0,t_0 \to n,t\}}  \!\!  P(\dtrajdiag) \, P_{n_0} 
(t_0) 
\  (\cdots)_{\dtrajdiag}\,, \qquad \ 
\label{Eq:traj-sum}
\end{eqnarray}
where $(\cdots)_{\dtrajdiag}$ is the quantity of interest for trajectory $\dtrajdiag$, and the sum is over all
trajectories from state $n_0$ (in the diagonal basis of the system's density matrix) at time $t_0$ 
to state $n$ (in the diagonal basis of the system's reduced density matrix) at time $t$.

The proof of the {\it integral fluctuation theorem} is carried out by considering the average,
\begin{align}
\left\langle \e^{-\Delta S_{\rm tot}} \right\rangle \ &= \  
\sum_{n_0,n} \ \sum_{\dtrajdiag \in  \{n_0,t_0 \to n,t\}}    P(\dtrajdiag) \, P_{n_0} 
(t_0) 
\ \e^{-\Delta S_{\rm tot}(\dtrajdiag)}\,,
\label{Eq:exponential-entropy-as-traj-sum}
\end{align}
where $ P(\dtrajdiag)$ is a trajectory in the set-up A defined in the paragraph above 
Eq.~(\ref{Eq:D_0,1,2-TR}).
Substituting in Eq.~(\ref{Eq:fluct-rel-ingredient}) on the right-hand-side gives
a result in terms of the trajectories in the time-reverse set-up (the one called set-up B above),
\begin{align}
\left\langle \e^{-\Delta S_{\rm tot}} \right\rangle \ &= \  
\sum_{n_0,n}\ \sum_{\dtrajdiag \in  \{n_0,t_0 \to n,t\}}    \overline{P}(\overline{\dtrajdiag}) \, P_{n} (t) \, .
\end{align}
The sum over all trajectories $\dtrajdiag$ from $n_0$ at time $t_0$ to $n$ at time $t$,
is replaced by a sum over all trajectories $\overline{\dtrajdiag}$ from $\overline{n}$ at time $t_0$ to $\overline{n_0}$ at time $t$ in set-up B, so
\begin{align}
\left\langle \e^{-\Delta S_{\rm tot}} \right\rangle \ &= \  
\sum_{n_0,n} \ \sum_{\overline{\dtrajdiag} \in  \{\overline{n},t_0 \to \overline{n_0},t\}}    \overline{P}(\overline{\dtrajdiag}) \, P_{n} (t)  \, .
\label{Eq:step-to-nepi}
\end{align}
Nothing changes if the sum over all $n_0$ is replaced by one over all $\overline{n_0}$.
The dynamics of the system in set-up B (whatever they may be) must conserve probability,
which means that the sum over all trajectories from $\overline{n}$ to $\overline{n_0}$ 
summed over all $\overline{n_0}$ must give unity;
\begin{eqnarray}
\sum_{\overline{n_0}}\ \sum_{\overline{\dtrajdiag} \in  \{\overline{n},t_0 \to \overline{n_0},t\}}    \overline{P}(\overline{\dtrajdiag}) \ =\ 1  \, .
\label{Eq:sum_overlineP_is_one}
\end{eqnarray}
This hold irrespective of the basis in which one writes the final state of the system, since probability conservation guarantees that the diagonal elements of a reduced density matrix sum to one in any basis.  This is convenient, because the final state of the evolution in set-up B (the sum over trajectories $\dtrajdiag$) is not usually diagonal in the basis used (which is the diagonal basis of the initial state of set-up A), as discussed at the end of section~\ref{Sect:Traj-TR}.

Substituting Eq.~(\ref{Eq:sum_overlineP_is_one}) into Eq.~(\ref{Eq:step-to-nepi}), the right hand side reduces to $\sum_n P_n(t)$,
this is a sum over the final state of the system in set-up A.  However, irrespective of the dynamics
of set-up A, conservation of probability tells us that $\sum_n P_n(t)=1$.  Thus we have proven the 
 {\it integral fluctuation theorem} in  Eq.~(\ref{Eq:non-equil_partition_identity}) under completely general conditions for an arbitrary quantum set-up described by any Hamiltonian of the form Eq.~(\ref{Eq:H_tot}) for any initial factorized state of system and reservoirs.

 The fact the proof is restricted to factorized states of system and reservoirs means it does not apply to situations in which the system is initially entangled with reservoir states.
 Below, in section~\ref{Sect:nonfactorized}, we will use the above proof as the principal ingredient in a proof of Eq.~(\ref{Eq:non-equil_partition_identity}) for arbitrary initial states including those where the system and reservoirs are initially entangled.
 
However, the above proof already applies to one of the most common experimental situations,
that where one has measured the system state at the beginning of the evolution in an arbitrary basis.  If the basis is not the system's energy eigenbasis then the system will be in a superposition of energy states, a situation which one cannot model with the classical rate equations 
 in Refs.~[\onlinecite{Schmiedl-Seifert2007,van-Broeck-review2015, our-review-2017}], irrespective of whether the dynamics are Markovian or not.
 
\subsection{Second law of thermodynamics}
\label{Sect:2ndlaw}

Since Eq.~(\ref{Eq:non-equil_partition_identity}) applies 
for any factorizable initial state,  it takes only one line of algebra \cite{Schmiedl-Seifert2007,van-Broeck-review2015, our-review-sect8.10} arrive at the  second law of thermodynamics on average
\begin{eqnarray}
\langle \Delta S_{\rm tot} \rangle  &\geq& 0\, .
\label{Eq:2ndlaw_on_average}
\end{eqnarray}
The proof is done by noting that $x \geq 1-\e^{-x}$ for all $x$ (this is easily seen graphically, but is formally an example of  Jensen's inequality), and so whatever the probability distribution of $\Delta S_{\rm tot}$, one must have 
$\langle \Delta S_{\rm tot} \rangle  \geq 1 -   \langle \e^{-\Delta S_{\rm tot}} \rangle =0$.

However, Eq.~(\ref{Eq:non-equil_partition_identity}) tells us more than this,
it tells us that all set-ups {\it must} sometimes have fluctuations in which $ \Delta S_{\rm tot} <0$.  
Hence, the second law is {\it only}  obeyed on average, 
and there will {\it always} be fluctuations (perhaps only very rare fluctuations) which violate it.
To see this, it is enough to note that if a set-up only had trajectories with   
$\Delta S_{\rm tot} (\dtrajdiag) > 0$ (positive entropy production), then it would have $ \left\langle \e^{-\Delta S_{\rm tot}} \right\rangle < 1$.   Thus, any set-up must also have trajectories with $\Delta S_{\rm tot} (\dtrajdiag) < 0$ to satisfy Eq.~(\ref{Eq:non-equil_partition_identity}).  The exponential factor in Eq.~(\ref{Eq:non-equil_partition_identity}) 
means that the probability of trajectories with $\Delta S_{\rm tot} (\dtrajdiag) < 0$ will be less than that of those with
$\Delta S_{\rm tot} (\dtrajdiag) >0$, but the probability of trajectories with  $\Delta S_{\rm tot} (\dtrajdiag) < 0$ cannot be zero.
The only exception to this statement, is a system in which no trajectories generate any entropy, so $\Delta S_{\rm tot} (\dtrajdiag) = 0$ for all $\dtrajdiag$.  
%This satisfies  Eq.~(\ref{Eq:non-equil_partition_identity}) and obviously has $\langle \Delta S_{\rm tot} \rangle = 0$.  It is 
% thermodynamically reversible, not just on average, but in every realization (as there are no fluctuations in entropy production). 
% A trivial example of this is when the quantum system is not coupled to the reservoirs.
% Other less trivial examples are difficult to find, because the entropy change is rarely strictly zero in any realistic system. 

\subsection{Jarzynski equality under certain conditions \label{Sect:Jarzynski}}

Let us consider the Jarzynski equality \cite{Jarzynski} generalized
to grand-canonical potentials.\cite{Schmiedl-Seifert2007}  It applies to a classical system that starts its evolution in thermal equilibrium 
at temperature $T$, that then experiences a time-dependent drive
and time-dependent coupling to multiple reservoirs at different chemical potentials, but all at temperature $T$.  This generalized Jarzynski equality
states that the work $\Delta W$ that is done on a system by the drive and the reservoirs obeys
\begin{eqnarray}
\left\langle e^{- \Delta W/T} \right\rangle &=& e^{-\Delta F/T} 
\label{Eq:Jarzynski}
\end{eqnarray}
where temperature is measured in units of energy, so $\kB=1$.
The free energy difference
\begin{eqnarray}
\Delta F= T \left(\ln[Z(\mu_0;t_0)] - \ln[Z(\mu_0;t)]\right) \, .
\label{Eq:DeltaF}
\end{eqnarray}
with $Z(\mu_0;\tau) = \e^{\mu_0/T} \sum_{n_0} \e^{E_{\rm sys}^{(n)}(\tau)/T}$.  
Here $Z(\mu_0,t_0)$ coincides with the partition function of the initial equilibrium state,
however the factors of 
$\e^{\mu_0/T} $ cancel in Eq.~(\ref{Eq:DeltaF}), so $\Delta F$ is independent of $\mu_0$.
The original Jarzynski equality is recovered in the limit where the system exchanges energy but not particles with the reservoirs.
The above generalized Jarzynski equality was proven for classical systems described by markovian rate equations in Refs.~[\onlinecite{Schmiedl-Seifert2007,van-Broeck-review2015}]. 

The derivation for non-Markovian quantum systems presented here is restricted to systems in which the system-reservoir coupling is reduced to zero at the end of the evolution at time $t$. This is in addition the assumption that system and reservoirs 
are in a product state (each in internal equilibrium at the same  temperature $T$).
In general, the system and reservoirs will arrive at time $t$ in a non-factorizable state, 
which will have a non-zero amount of energy in the system-reservoir coupling.  Turning off the system-reservoir coupling (typically by changing the voltage on the gate that separates the system from the reservoirs) will thus change the set-up's energy, and thus corresponds to work being done by the drive.  We include this work done to turn off the system-environment coupling  in $W$ in Eq.~(\ref{Eq:Jarzynski}).

Let us be clear, this restriction is a way of {\it avoiding} the problem of the energy in the system-reservoir coupling, by having it be zero at the beginning and end of the evolution.  This is a small step beyond the proof for Markovian classical systems in Refs.~[\onlinecite{Schmiedl-Seifert2007,van-Broeck-review2015}], because the dynamics can be non-Markovian between time $t_0$ and time $t$ (as well of course allowing for quantum physics).  However, it is hoped that future work might reveal a more general Jarzynski equality, that holds when the energy in the system-reservoir coupling is non-zero at the beginning or end of the evolution.  One possible direction for this future work is to compare with the situation where all reservoir chemical potentials are equal (so the reservoirs do no work on the system), for which there is an elegant proof of the Jarzynski equality in Ref.~[\onlinecite{Campisi-review2011}].

The proof presented here makes use of Eq.~(\ref{Eq:central-obs}), but involves 
different rotations at the beginning and end of the evolution from those discussed
below Eq.~(\ref{Eq:central-obs}).  Instead of rotations to the basis
where the system's density matrix is diagonal, one rotates to the basis in
which the system's Hamiltonian $H_{\rm sys}$ is diagonal.  
Thus ${\cal W}_0$ is the rotation from the diagonal basis of $H_{\rm sys}(t_0)$ to the basis in which the evolution is calculated (if these bases are the same, then ${\cal W}_0=1$). Similarly, ${\cal W}$ is the rotation from the basis in which the evolution is calculate to the basis in which $H_{\rm sys}(t)$ is diagonal.
While these rotations are different from those below Eq.~(\ref{Eq:central-obs}), they are still unitary, which means they do not affect the trajectory's entropy, hence 
Eq.(\ref{Eq:central-obs-diag}) still holds.

Consider a trajectory $\dtrajdiag$ from system state $n_0$ at time $t_0$ to system state $n$ in time $t$,
where $n_0$ is the eigenstate of $H_{\rm sys}(t_0)$ with energy $E_{n_0}(t_0)$
and  $n$ is the eigenstate of $H_{\rm sys}(t)$ with energy $E_{n}(t)$.
Then the work done on the system by the driving is 
\begin{eqnarray}
\Delta W_{\rm drive}(\dtrajdiag) = \big[E_{\rm sys}^{(n)}(t) -  E_{\rm sys}^{(n_0)}(t_0)\big] +\sum_\alpha  \Delta E_\alpha (\dtrajdiag)\, . \ \ 
\label{Eq:W_drive}
\end{eqnarray}
The square brackets is the work done by the drive which stays in the system, 
while $\Delta E_\alpha (\dtrajdiag)$ is defined as the energy flow into reservoir $\alpha$ during the trajectory $\dtrajdiag$. Note that this equality holds because of the above restriction to systems in which there is no energy in the system-reservoir coupling at the beginning or end of the evolution.
The work done on the system by the reservoirs during trajectory $\dtrajdiag$ is given by
\begin{eqnarray}
\Delta W_{\rm res}(\dtrajdiag)  =- \sum_\alpha  \mu_\alpha \,\Delta N_\alpha (\dtrajdiag) \, ,
\end{eqnarray}
where $\Delta N_\alpha (\dtrajdiag)$ is the number of particles flowing into reservoir $\alpha$ during the trajectory $\dtrajdiag$.
Given Eq.~(\ref{Eq:deltaS_alphak}), one sees that 
\begin{eqnarray}
\Delta S_{\rm res}(\dtrajdiag) = {1 \over T}\sum_{\alpha} \big(\Delta E_\alpha (\dtrajdiag)- \mu_\alpha \,\Delta N_\alpha (\dtrajdiag)\big),
\end{eqnarray}
since all reservoirs have the same temperature.  Thus,
\begin{eqnarray}
\Delta W(\dtrajdiag) &=&  
E_{\rm sys}^{(n)}(t) -  E_{\rm sys}^{(n_0)}(t_0) + T \, \Delta S_{\rm res}(\dtrajdiag) \, , \ \ 
\label{Eq:Jarzynski-DS_res}
\end{eqnarray}
where $\Delta W(\dtrajdiag) =\Delta W_{\rm drive}(\dtrajdiag) +\Delta W_{\rm res}(\dtrajdiag) $ is the total work done on the system.  
Using this equality in the average of $\exp[-\Delta W(\dtrajdiag)/T]$  over all $\dtrajdiag$, defined in Eq.~(\ref{Eq:traj-sum}),
\begin{eqnarray}
\big\langle e^{- \Delta W/T} \big\rangle &=&
\sum_{n_0,n} \ 
\sum_{\dtrajdiag \in  \{n_0,t_0 \to n,t\}}  \!\!  P(\dtrajdiag) \e^{-\Delta S_{\rm res}(\dtrajdiag)} 
\nonumber \\
& & \times
\e^{-[E_{\rm sys}^{(n)}(t) -  E_{\rm sys}^{(n_0)}(t_0)]/T}  P_{n_0} (t_0) \,, \qquad \ 
\label{Eq:Jarzynski_step-one}
\end{eqnarray}
Eq.~(\ref{Eq:central-obs}) is now used to write this in terms of $\overline{P}(\overline{\dtrajdiag})$.
In other words, the average over trajectories in a set-up A is written in terms of the trajectories in set-up B (defined earlier as the time-reverse of set-up A).
The initial system density matrix (at time $t_0$)  is diagonal 
in the eigenbasis of $H_{\rm sys}(t_0)$, and the probability of being in state $n_0$ is
\begin{eqnarray}
P_{n_0}(t_0) &=&{1 \over Z_0(\mu_0)}  
\ \e^{-\big(E_{\rm sys}^{(n_0)}(t_0) -\mu_0\big)/T}  ,\qquad
%\exp\big[-\big(E_{\rm sys}^{(n_0)}(t_0) -\mu_0\big) \big/T\big] \, ,\qquad
\label{Eq:Jarzynski_Pinitial}
\end{eqnarray}
with $Z_0(\mu_0)$ given below Eq.~(\ref{Eq:Jarzynski}).
Then Eq.~(\ref{Eq:Jarzynski_step-one}) becomes
\begin{eqnarray}
\big\langle e^{- \Delta W/T} \big\rangle &=&{1 \over Z_0(\mu_0)}  
\sum_{n_0,n} \e^{-[E_{\rm sys}^{(n)}(t)-\mu_0]/T}
\nonumber \\
& & \qquad \times
\sum_{\overline{\dtrajdiag} \in  \{\overline{n},t_0 \to \overline{n_0},t\}}  \!\!  \overline{P}(\overline{\dtrajdiag}) \, ,
\label{Eq:Jarzynski_step-two}
\end{eqnarray}
where the sum over all $\dtrajdiag$ from $n_0$ to $n$ in set-up A,
has become a sum over all $\overline{\dtrajdiag}$ from $\overline{n}$ to $\overline{n_0}$ in set-up B.
Nothing changes if the sum over all $n_0$ is replaced by one over all $\overline{n_0}$.
Irrespective of the dynamics in set-up B, 
the sum over all trajectories with final state $\overline{n_0}$, summed over all $\overline{n_0}$ must give one.
Therefore Eq.~(\ref{Eq:Jarzynski_step-two}) reduces to 
$\left\langle e^{- \Delta W/T} \right\rangle = Z(\mu_0)/Z_0(\mu_0)$. 
Now using Eq.~(\ref{Eq:DeltaF}), one immediately gets the generalized Jarzynski equality in Eq.~(\ref{Eq:Jarzynski}).

One can apply Jensen's inequality to Eq.~(\ref{Eq:Jarzynski}) to find a well-known formulation of the second law, 
\begin{eqnarray}
\langle \Delta W \rangle &\leq& \Delta F
\end{eqnarray}
but the assumptions and restrictions in this derivation to Eq.~(\ref{Eq:Jarzynski}) make this a less general version
of the second law than that in Eq.~(\ref{Eq:2ndlaw_on_average}).

\subsection{Crooks equation}
\label{Sect:Crooks}

The Crooks equation\cite{Crooks1999} is a relation between the dynamics of a set-up A and a set-up B (which has the time-reverse of set-up A) in situations where the system undergoes time-dependent driving while coupled to reservoirs.
Consider set-up A described by the Hamiltonian in Eq.~(\ref{Eq:H_tot}) starting at time $t_0$ with the system's density matrix $\rho_{\rm sys}^{\rm (i)}$ (``i'' for initial), and ending the evolution at time $t$ with the systems reduced density matrix being $\rho_{\rm sys}^{\rm (f)}$ (``f'' for final).
Let us define $P\big(\Delta S_{\rm tot};\rho_{\rm sys}^{\rm (i)}\to \rho_{\rm sys}^{\rm (f)}\big)$ as the probability that set-up A would have a total entropy changes of $\Delta S_{\rm tot}$ between $t_0$ and $t$.
Now let us consider set-up B described by the time-reverse of  Eq.~(\ref{Eq:H_tot}), and take its initial system density matrix to be $\overline{\rho}_{\rm sys}^{\rm (f)}  \equiv \Theta_{\rm sys}^\dagger \rho_{\rm sys}^{\rm (f)} \Theta_{\rm sys}$ where $\Theta_{\rm sys}$ is the time-reversal operator on the system alone; so set-up B's  initial system state  is the time-reverse of 
set-up A's final system state.
Set-up B's evolution will not be the time-reverse of set-up A's, because we do not time-reverse individual reservoir states, cf.\ section~\ref{Sect:non-reversal_of_res_states}.
Let its evolution under the time-reverse of  Eq.~(\ref{Eq:H_tot}), so that its reduced system density matrix at time $t$ is  $\overline{\rho}_{\rm sys}^{\rm (f2)}$.
Let us then define 
$\overline{P}\big(\Delta S_{\rm tot}; \overline{\rho}_{\rm sys}^{\rm (f)} \to \overline{\rho}_{\rm sys}^{\rm (f2)}\big)$ 
as the probability that set-up B would have a total entropy changes of $\Delta S_{\rm tot}$ between $t_0$ and $t$.
Below we will prove the following slight generalization of the Crooks equation
for non-Markovian quantum systems, it reads
\begin{eqnarray}
& &\overline{P}\big(-\Delta S_{\rm tot};\, \overline{\rho}_{\rm sys}^{\rm (f)} \to  \overline\rho_{\rm sys}^{\rm (f2)}\big) 
\nonumber \\
& & \qquad \qquad
= P\big(\Delta S_{\rm tot};\,\rho_{\rm sys}^{\rm (i)} \to ,\rho_{\rm sys}^{\rm (f)}\big)\, \e^{-\Delta S_{\rm tot}} , \qquad
\label{Eq:Crooks}
\end{eqnarray}
{\it under the condition that the time-reverse of the final state of the evolution in set-up B\dRed{\sout{, $\rho_{\rm sys}^{\rm (f2)}$,}} is \dRed{\sout{diagonal in the same basis} the same} as the initial state in set-up A,}  
\dRed{i.e. $\rho_{\rm sys}^{\rm (f2)}=\rho_{\rm sys}^{\rm (i)}$.
\sout{This is slightly more general than the condition used by Crooks for classical systems \cite{Crooks1999},
since his condition was that the final state in set-up B was the same as (the time-reverse\cite{footnote:Crooks} of) the initial state in set-up A.}}

%%%%%%%%%%%%%%%%%%%%%%%%%%%%
\begin{figure*}
\includegraphics[width=0.78\textwidth]{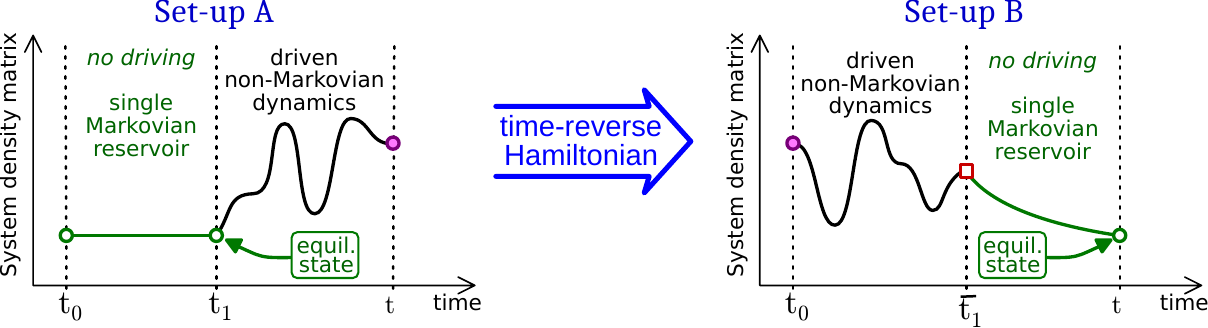}
\caption{\label{Fig:Crooks}
A sketch of a situation for which the Crooks equation in Eq.~(\ref{Eq:Crooks}) is applicable.
The plots show a {\it cartoon} of how the system density matrix (vertical axis) 
varies with time from $t_0$ to $t$.
In set-up A, the system starts in a thermal state (open green circle). For time $t_0$ to time $t_1$, the system 
has a coupling to a single Markovian reservoir with which it is in equilibrium.  Thus at time $t_1$
the system is still in the thermal state. For time $t_1$ to time $t$, the system is driven while interacting with many non-Markovian reservoirs, so that at time $t$ it is in a highly non-equilibrium state (filled purple circle).
In set-up B the system starts in the time-reverse of set-up A's final state, and evolves under the time-reversed Hamiltonian.  Between $t_0$ and $\overline{t_1}=t+t_0-t_1$ it undergoes driven non-markovian dynamics,
so it is in some non-equilibrium state (red square) at time $\overline{t_1}$.  However, after that it is only coupled to a single Markovian reservoir, so it decays towards the thermal state in equilibrium with that reservoir.
We assume it reaches that state at time $t$.  Thus the initial state in both set-ups is the final state in the other,
which is sufficient that Eq.~(\ref{Eq:Crooks}) is applicable,
even though the evolution in set-up A is completely arbitrary and non-Markovian between $t_1$ and $t$.
Finally, as nothing happens to the state in set-up A between $t_0$ and $t_1$,  we can equally find a Crooks equation of the type in Eq.~(\ref{Eq:Crooks}) between the dynamics from time $t_1$ to time $t$ in set-up A and the dynamics
from time $t_0$ to time $t$ in set-up B.
}
\end{figure*}
%%%%%%%%%%%%%%%%%%%%%%%%%%

In general, there is no reason to expect the condition below Eq.~(\ref{Eq:Crooks}) to hold;  if set-up B's initial state is the time-reverse of the final state of set-up A, it is likely to end up in some state $\overline{\rho}_{\rm sys}^{\rm (f2)}$, whose time-reverse $\rho_{\rm sys}^{\rm (f2)}$ has nothing to do with $\rho_{\rm sys}^{\rm (i)}$. 
Thus, in general Eq.~(\ref{Eq:Crooks}) will not be satisfied, but there are scenarios of interest in which the condition is satisfied. 
Fig.~\ref{Fig:Crooks} shows a situation (a quantum version of a scenario proposed by Crooks \cite{Crooks1999}) 
in which one naturally has  $\rho_{\rm sys}^{\rm (f2)}= \rho_{\rm sys}^{\rm (i)}$.

\dRed{}

\subsubsection{Proof of the Crooks equation}

To derive Eq.~(\ref{Eq:Crooks}) from  Eq.~(\ref{Eq:central-obs}) we can follow
the proof in Ref.~[\onlinecite{van-Broeck-review2015}]. 
The probability that the entropy change is $\Delta S_{\rm tot}$ in the time from $t_0$ to $t$ is
\begin{eqnarray}
P\big(\Delta S_{\rm tot};\rho_{\rm sys}^{\rm (i)}\to \rho_{\rm sys}^{\rm (f)}\big) \!&=&\! \sum_{n_0,n} \ \sum_{\dtrajdiag \in \{n_0,t_0 \to n,t\}} P(\dtrajdiag) P_{n_0}^{\rm (i)}
\nonumber \\
& & \qquad  \times 
\delta \big[\Delta S_{\rm tot}(\dtrajdiag)-\Delta S_{\rm tot}\big], \qquad
\end{eqnarray}
where the $\delta$-function picks out only those trajectories with entropy change $\Delta S_{\rm tot}$,
and
$P_{n_0}^{\rm (i)}$ is the $n_0$th element of $\rho_{\rm sys}^{\rm (i)}$ in its diagonal basis. 
The $\delta$-function means that the equality holds if one multiples the left hand side by $\e^{-\Delta S_{\rm tot}}$ and the right hand side by $\e^{-\Delta S_{\rm tot}(\dtrajdiag)}$.
Eq.~(\ref{Eq:fluct-rel-ingredient}) 
--- derived above from Eq.~(\ref{Eq:central-obs}) --- 
can be used to write
\begin{eqnarray}
& & \hskip -10mm P\big(\Delta S_{\rm tot};\rho_{\rm sys}^{\rm (i)}\to \rho_{\rm sys}^{\rm (f)}\big) \,\e^{-\Delta S_{\rm tot}} 
\nonumber \\
\!&=&\! \sum_{n_0,n} \ \sum_{\dtrajdiag \in \{n_0,t_0 \to n,t\}} \overline{P}(\overline{\dtrajdiag}) P_{n}^{\rm (f)} 
%\nonumber \\
%& & \qquad \qquad \times 
\delta [\Delta S_{\rm tot}(\dtraj)-\Delta S_{\rm tot}]. 
\nonumber
\end{eqnarray}
This means the dynamics are now written in terms of trajectories in the time-reversed set-up (set-up B).
Rewriting the sum over $\dtrajdiag$ from $n_0$ to $n$ as a sum over $\overline{\dtrajdiag}$ from $\overline{n}$ to $\overline{n_0}$, and using the fact that if \dRed{$\rho_{\rm sys}^{\rm (f2)}=\rho_{\rm sys}^{\rm (i)}$} then $\overline{\Delta S_{\rm tot}}(\overline{\dtraj}) = -\Delta S_{\rm tot}(\dtraj)$, leads to
\begin{eqnarray}
& & \hskip -10mm
P\big(\Delta S_{\rm tot};\rho_{\rm sys}^{\rm (i)}\to \rho_{\rm sys}^{\rm (f)}\big) \,\e^{-\Delta S_{\rm tot}} 
\nonumber \\
\!&=&\! \sum_{\overline{n_0},\overline{n}} \ \sum_{\overline{\dtrajdiag} \in \{\overline{n},t_0 \to \overline{n_0},t\}} \!\overline{P}(\overline{\dtrajdiag}) P_{n}^{\rm (f)}
%\nonumber \\
%& & \quad \times
\, \delta [\Delta S_{\rm tot}(\dtraj)+\Delta S_{\rm tot}],
\nonumber \\
\label{Eq:Crooks-step1}
\end{eqnarray}
where we have used the fact that nothing changes when the sum over $n_0$ and $n$
is replaced by a sum over $\overline{n_0}$ and $\overline{n}$. 
Recalling that $P_n^{\rm (f)}$ is the probability that set-up A finishes in state $n$ 
in the diagonal basis of $\rho_{\rm sys}^{\rm (f)}$, 
one can always choose the initial density matrix in set-up B to be  $\overline{\rho}_{\rm sys}^{\rm (f)}$.  
Then, the probability that the system starts in state $\overline{n}$ in the diagonal basis of  $\overline{\rho}_{\rm sys}^{\rm (f)}$
equals $P_n^{\rm (f)}$. Hence, it looks like the right hand side of Eq.~(\ref{Eq:Crooks-step1}) 
equals $\overline{P}\big(-\Delta S_{\rm tot};\, \overline{\rho}_{\rm sys}^{\rm (f)}\to  \overline{\rho}_{\rm sys}^{\rm (f2)}\big)$, 
whatever final system density matrix, $ \overline{\rho}_{\rm sys}^{\rm (f2)}$, this evolution may give.
However, this is overlooking the fact that the trajectories $\overline{\dtrajdiag}$ end in the diagonal basis of 
 $\overline{\rho}_{\rm sys}^{\rm (i)}$, thus the right hand side of  Eq.~(\ref{Eq:Crooks-step1}) 
 only equals $\overline{P}\big(-\Delta S_{\rm tot}\,;\, \overline{\rho}_{\rm sys}^{\rm (f)}\to \overline{\rho}_{\rm sys}^{\rm (f2)}\big)$, if $\rho_{\rm sys}^{\rm (f2)}$ is diagonal in the same basis as $\rho_{\rm sys}^{\rm (i)}$.
\dRed{Furthermore, we used $\overline{\Delta S_{\rm tot}}(\overline{\dtraj}) = -\Delta S_{\rm tot}(\dtraj)$ above Eq.~(\ref{Eq:Crooks-step1}), and this is not true unless the dynamics obey the condition $\rho_{\rm sys}^{\rm (f2)}=\rho_{\rm sys}^{\rm (i)}$;
the published version of this article overlooked this condition, and so made an erroneous statement about the condition of validity of the Crooks equation
which has been corrected here\cite{footnote:error}.}
So one only recovers Eq.~(\ref{Eq:Crooks}), 
if the dynamics satisfy the condition below Eq.~(\ref{Eq:Crooks}).

%===================================
\section{Fluctuation theorems for non-factorizable initial conditions}
\label{Sect:nonfactorized}

The system and reservoirs can either be in a factorizable  or  non-factorizable state;
a factorizable state being one where the total density matrix can be written as a product of the system density matrix and the reservoir density matrices; e.g..~$\hat\rho_{\rm sys} \otimes \hat{\rho}_{\rm res 1} \otimes  \hat{\rho}_{\rm res 2} \otimes \cdots$.  This state is often also called a product state.   A non-factorizable state is any other density matrix for the system plus the reservoirs.
Up to this point, this work has discussed a set-up 
which started its evolution at time $t_0$ in a factorized state.  
In this section, we consider protocols in which the initial density matrix is in a non-factorizable state.

In quantum mechanical systems, a system's state is changed by the mere fact of observing it.
In particular, the act of measuring the system state projects it into a definite system state, which means the entanglement with the reservoirs is destroyed, leaving the system and reservoirs in a factorized density matrix. 
Hence,
the only way to measure the changes between time $t_1$ and time $t_2$ without this projection onto a factorized state at time $t_1$ is to consider the following protocol.
\begin{itemize}
\item[] {\bf Non-factorizing Protocol.} 
We prepare many set-ups in the same manner (starting each with the same factorized state at a time $t_0$ and letting them evolve in the same manner), so they are all in the same non-factorized state at a time $t_1$.
We then split them into two groups (i and ii).  We measure group i at time $t_1$ and we measure group ii at a later time $t_2$.  
As we do not measure the set-ups in group ii at time $t_1$, they are not projected on to a factorized state
at time $t_1$.  Despite this, we know about the state of the system at time $t_1$ from the measurements on
set-ups in group i.
This enables us to see the difference between the set-up's properties at time $t_2$, and its properties at  
the earlier time $t_1$, when it was in a non-factorized state at time $t_1$.
\end{itemize}
It is important to note that this protocol cannot be used to study correlations between the state at time $t_1$ and time $t_2$, because one measurement is on group i and the other is on group ii.
For example, we can see how the distribution of entropy changes between time $t_1$ and time $t_2$,
but we cannot see how a fluctuation of entropy (say the system having much less entropy than average) 
at time $t_1$ correlates with a fluctuation of entropy at the later time $t_2$.

\subsection{Conditional probability in this  protocol}

Let us consider the above non-factorizing protocol being used to study the changes 
in a set-up between time $t_1$ and time $t_2$, when the system is in a non-factorized state at time $t_1$. 
For this one can assume the set-up was prepared in the distant past at time $t_0$ in a factorized state, but that  the system has interacted with the reservoirs for so long by the time $t_1$ that 
it is in a highly complicated entangled state with the reservoirs.  Our main interest is in 
situations where the time $t_0$ was so far in the past, that the dynamics at the times of interest ($t_1$ and $t_2$) {\it do not depend} on the choice of the system state at time $t_0$.

Consider $P(\Delta S^{\rm tot}_1; t_1,t_0)$ to be the probability distribution of the entropy change $\Delta S^{\rm tot}_1$ between the time in the distant past $t_0$ and time $t_1$, as measured on set-ups in group i.  
Then consider $P(\Delta S^{\rm tot}_2; t_2,t_0)$ to be the probability distribution of the entropy change $\Delta S^{\rm tot}_2$ between the time in the distant past $t_0$ and time $t_2$,  as measured on set-ups in group ii.  
Then one can define $Q(\Delta S^{\rm tot}_{2\leftarrow1}; t_2,t_1)$ as a conditional probability distribution, 
for the entropy change of 
\begin{eqnarray}
\Delta S^{\rm tot}_{2\leftarrow1} =\Delta S^{\rm tot}_2 -\Delta S^{\rm tot}_1\, ,
\label{Eq:S_1_to_2}
\end{eqnarray}
between time $t_1$ and time $t_2$.
This means that $Q(\Delta S^{\rm tot}_{2\leftarrow1}; t_2,t_1)$ measures how the probability distribution changes
between $t_1$ and $t_2$.
It obeys
\begin{eqnarray}
P\big(\Delta S^{\rm tot}_2;t_2,t_0) &=& \int \rmd (\Delta S^{\rm tot}_1) \ Q\big(\Delta S^{\rm tot}_2 -\Delta S^{\rm tot}_1; t_2,t_1\big) 
\nonumber \\
& & \qquad \qquad \times P\big(\Delta S^{\rm tot}_1; t_1,t_0\big). \qquad 
\label{Eq:Def-Q}
\end{eqnarray}

One can always define the function $Q(\Delta S^{\rm tot}_{2\leftarrow1}; t_2,t_1)$ in this manner. 
However, the price to pay for highly non-Markovian dynamics (strong memory effects),
is that it may depend on both the initial state of the set-up at time $t_0$, 
and on the dynamics of the set-up from time $t_0$ to time $t_1$ (as well as the dynamics from
$t_1$ to $t_2$).
Thus this is not a pleasant quantity to consider in general.  
However, it becomes much more natural in situations where $t_0$ is far enough in the past that
 $Q(\Delta S^{\rm tot}_{2\leftarrow1}; t_2,t_1)$ depends weakly on it, and where the dynamics for a long time
 before time $t_1$ are simple enough to treat in some manner.
 An ideal example, which we will consider in more detail below is when the 
 system Hamiltonian is time-independent for a long enough time before $t_1$ that the set-up
 has achieved a {\it steady-state} at time $t_1$.

\subsection{Integral fluctuation theorem}
\label{Sect:NEPI-2}

Now we use the proof of the integral fluctuation theorem, Eq.~(\ref{Eq:non-equil_partition_identity}),
for factorized initial conditions, to prove that it also holds for the entropy change between time $t_1$ and $t_2$,
when the set-up is in an arbitrary non-factorized state at both $t_1$ and $t_2$.
In this context, we assume the entropy change is measured via the non-factorizing protocol above,
in which the set-up  was in a factorized state at a time $t_0$ in the distant past (long before the times of interest, $t_1$ and $t_2$).

As above we assume $\Delta S^{\rm tot}_2$ is the entropy change from time $t_0$ to time $t_2$ 
(as measured on set-ups in group ii of the non-factorizing protocol). Then  
Eq.~(\ref{Eq:non-equil_partition_identity}), 
proven for a factorized state at time $t_0$ in section~\ref{Sect:NEPI}  above, becomes
\begin{eqnarray}
\big\langle e^{-\Delta S^{\rm tot}_2} \big\rangle
\ = \, \int \rmd (\Delta S^{\rm tot}_2) P(\Delta S^{\rm tot}_2; t_2,t_0) e^{-\Delta S^{\rm tot}_2}. \qquad \ 
\end{eqnarray}
Substituting in Eq.~(\ref{Eq:Def-Q}),  
\begin{eqnarray}
\big\langle e^{-\Delta S^{\rm tot}_2} \big\rangle
&=& \int \rmd (\Delta S^{\rm tot}_{2\leftarrow1})\,\rmd (\Delta S^{\rm tot}_1) \  Q(\Delta S^{\rm tot}_{2\leftarrow1}; t_2,t_1)
\nonumber \\
& & \qquad \times P(\Delta S^{\rm tot}_1; t_1,t_0) \e^{-\Delta S^{\rm tot}_{2\leftarrow1}-\Delta S^{\rm tot}_1}
\nonumber \\
&=& 
\int \rmd (\Delta S^{\rm tot}_{2\leftarrow1}) \, Q(\Delta S^{\rm tot}_{2\leftarrow1}; t_2,t_1)e^{-\Delta S^{\rm tot}_{2\leftarrow1}} 
\nonumber \\
& & \qquad \qquad \qquad \times \big\langle e^{-\Delta S^{\rm tot}_1} \big\rangle 
\end{eqnarray}
where $\left\langle \exp[-\Delta S^{\rm tot}_1] \right\rangle$ is the average over dynamics from time $t_0$ to $t_1$.
Now substituting in Eq.~(\ref{Eq:non-equil_partition_identity}) for the two averages, we have
\begin{eqnarray}
1 &=& \int \rmd (\Delta S^{\rm tot}_{2\leftarrow1}) \ Q(\Delta S^{\rm tot}_{2\leftarrow1}; t_2,t_1)\ e^{-\Delta S^{\rm tot}_{2\leftarrow1}} 
\nonumber \\
& &\equiv \big\langle e^{-\Delta S^{\rm tot}_{2\leftarrow1}}\big\rangle \ ,
\label{Eq:non-equil-partition-identity-non-factorized}
\end{eqnarray}
where $\Delta S^{\rm tot}_{2\leftarrow1}$ is the entropy change in the set-up between time $t_1$ and $t_2$.
Hence, we have the integral fluctuation theorem in Eq.~(\ref{Eq:non-equil_partition_identity}) 
for any non-factorized initial state.
The initial state now being the time at which one starts to study the set-up  (time $t_1$).

The average, $\langle \cdots \rangle$, in Eq.~(\ref{Eq:non-equil-partition-identity-non-factorized}) 
is defined via the non-factorizing protocol above, which relates changes to the
the difference between the set-up's non-factorised state at time $t_1$ 
(as measured on set-ups in group i of the non-factorizing protocol), 
and the set-up's non-factorised state at a later time $t_2$  
(as measured on set-ups in group ii of the non-factorizing protocol).

It immediately follows from this proof, that all statements about the second law of thermodynamics 
in Section~\ref{Sect:2ndlaw} above also hold for non-factorized states. 
The second law is always true on average, 
\begin{eqnarray}
\big\langle \Delta S^{\rm tot}_{2\leftarrow1}\big\rangle\geq 0,
\end{eqnarray}
irrespective of whether the set-up is in a factorized state at time $t_1$ or not.
Hence the average entropy will never be smaller at time $t_2$ than at time $t_1$ (for any $t_2 > t_1$), 
However, there {\it must} also be fluctuations for which $\Delta S^{\rm tot}_{2\leftarrow1} < 0$, if Eq.~(\ref{Eq:non-equil-partition-identity-non-factorized}) is to be satisfied.

\subsection{Steady-state fluctuation relation}

We can expect that a large class of non-Markovian systems will decay to a situation of steady state flow,
if the system is coupled to two or more reservoirs at different 
temperatures and electro-chemical potentials, while the Hamiltonian is kept time-independent.
Here we consider the case where there is a 
single steady-state (for a given Hamiltonian and given reservoir parameters), 
 which {\it all  initial states decay to}.
This steady-state will generally not be a factorizable state of the system and the reservoirs, since  the system will be entangled with at least some reservoir modes at all times.

The objective here is to derive the Evans-Searles fluctuation relation \cite{Evans-Searles1994}  for such a non-Markovian system for which the steady-state is non-factorizable.
For this, consider the non-factorizing protocol above, in which time $t_0$ is so far in the past that the choice of initial state at $t_0$ is 
irrelevant for the steady-state dynamics at the times of interest ($t_1$ and $t_2$).
Since one is completely free to choose the system state at time $t_0$, take it to coincide with that given by the steady-state when one traces out the reservoirs.   Then, by construction, the initial system density-matrix 
and the reduced final system density matrix are the same. This means the set-up obeys the Crooks equality derived above in section~\ref{Sect:Crooks}.
We also assume the Hamiltonian is invariant under time-reversal, 
such as is the case if it is time-independent, and has no external magnetic field. 
This means that the dynamics in the time-reversed set-up are the same as in the original set-up.
Then the Crooks equality 
for the entropy change between time $t_0$ and time $t_2$ reads
\begin{eqnarray}
P(-\Delta S^{\rm tot}_2;t_2,t_0) = P(\Delta S^{\rm tot}_2;t_2,t_0)e^{-\Delta S^{\rm tot}_2}
\label{Eq:Crooks-time-reversal-symm}
\end{eqnarray}
where we have dropped the overline on the left, because time-reversal changes nothing.
Now  Eq.~(\ref{Eq:Def-Q}) is used
to write the right hand side as evolution from time $t_0$ to time $t_1$ followed by evolution from
$t_1$ to $t_2$, as follows
\begin{eqnarray}
& & \hskip -4mm P(\Delta S^{\rm tot}_2;t_2,t_0)  e^{-\Delta S^{\rm tot}_2}
\nonumber \\
&=& \e^{-\Delta S^{\rm tot}_2} \int \rmd  (\Delta S^{\rm tot}_1) \ Q\big(\Delta S^{\rm tot}_2-\Delta S^{\rm tot}_1;t_2,t_1\big) 
\nonumber \\
& & \qquad \qquad \qquad \times P(\Delta S^{\rm tot}_1;t_1,t_0) \, .
\label{Eq:evans-searles-step1}
\end{eqnarray}
By the same logic the left hand side of Eq.~(\ref{Eq:Crooks-time-reversal-symm}) is
\begin{eqnarray}
& & \hskip -8mm
P\big(-\Delta S^{\rm tot}_2;t_2,t_0\big) 
\nonumber \\
&=& \int \rmd  (\Delta S^{\rm tot}_1) \ Q\big(\Delta S^{\rm tot}_1-\Delta S^{\rm tot}_2;t_2,t_1\big) 
\nonumber \\
& & \qquad \qquad \qquad \times
P\big(-\Delta S^{\rm tot}_1;t_1,t_0\big) \quad
\nonumber \\
&=& \e^{-\Delta S^{\rm tot}_2} \int \rmd  (\Delta S^{\rm tot}_1) \ Q\big(\Delta S^{\rm tot}_1-\Delta S^{\rm tot}_2;t_2,t_1\big)
\nonumber \\
& & \qquad \qquad \qquad \times \e^{\Delta S^{\rm tot}_2-\Delta S^{\rm tot}_1} P\big(\Delta S^{\rm tot}_1;t_1,t_0\big)   \, , \qquad
\label{Eq:evans-searles-step2}
\end{eqnarray}
where the last line comes from substituting in the Crooks equality, as applied to 
the evolution from time $t_0$ to $t_1$, for which it takes the form
$P(-\Delta S^{\rm tot}_1;t_1,t_0) =  P(\Delta S^{\rm tot}_1;t_1,t_0)e^{-\Delta S^{\rm tot}_1}$.

Note that the integrals in Eq.~(\ref{Eq:evans-searles-step1}) and Eq.~(\ref{Eq:evans-searles-step2})
are both convolutions of $P(\Delta S^{\rm tot}_1;t_1,t_0)$ with another function, in the former case
that function is $Q(\Delta S^{\rm tot}_2-\Delta S^{\rm tot}_1;t_2,t_1)$ and in the latter case that function is 
$Q(-\Delta S^{\rm tot}_2+\Delta S^{\rm tot}_1;t_2,t_1)e^{\Delta S^{\rm tot}_2 -\Delta S^{\rm tot}_1}$.  
Substituting  Eq.~(\ref{Eq:evans-searles-step1}) and Eq.~(\ref{Eq:evans-searles-step2}) into the right and left hand sides of Eq.~(\ref{Eq:Crooks-time-reversal-symm}), gives us an equality between the two convolutions,
\begin{eqnarray}
 & & \hskip -5mm \int \rmd  (\Delta S^{\rm tot}_1) \, Q(\Delta S^{\rm tot}_2-\Delta S^{\rm tot}_1;t_2,t_1) \,P(\Delta S^{\rm tot}_1;t_1,t_0)
 \nonumber \\
 &=&\!\! \int \rmd  (\Delta S^{\rm tot}_1) \, Q\big(\Delta S^{\rm tot}_1-\Delta S^{\rm tot}_2;t_2,t_1\big) \e^{\Delta S^{\rm tot}_2-\Delta S^{\rm tot}_1}\,
\nonumber \\
& & \qquad \qquad \qquad \times 
P\big(\Delta S^{\rm tot}_1;t_1,t_0\big) .
\label{Eq:evans-searles-step3}
\end{eqnarray}
This equality has the mathematical structure
\begin{eqnarray}
\int \rmd x A_1(y-x)B(x) = \int \rmd x A_2(y-x)B(x) \quad \mbox{ for all } y. 
\nonumber
\end{eqnarray}
We wish to show that this implies that the functions $A_1(x)$ and $A_2(x)$ are identical,
irrespective of the form of  $A_1(x)$ and $B(x)$.
To do this we consider the Fourier transforms of the functions defined as 
$A_i(x)=\int \rmd k \,a_i(k)\,\e^{\rmi kx}$ for $i=1,2$, and $B(x)=\int \rmd k \,b(k)\,\e^{\rmi kx}$.
We assume that the functions  $a_1(k)$, $a_2(k)$ and $b(k)$ are  well-behaved,
and also assume that  $b(k)$ is not zero over any finite range of $k$.
Given the Fourier transforms we have
\begin{eqnarray}
\int \rmd y \e^{-\rmi ky} \int\rmd x \, A_1(y-x)B(x)
\, =\,(2\pi)^2 a_1(k)b(k). \quad
\end{eqnarray}
with a similar equation for $A_2$ in place of $A_1$. This immediately gives
$a_2(k)=a_1(k)$ for all $k$ where $b(k)\neq 0$ (however it tells us nothing about the relationship between $a_2(k)$ and $a_1(k)$ when $b(k)=0$).   So long as $b(k)$ is not zero over a finite range of $k$, then it is sufficient to perform the inverse Fourier transform on $a_1(k)$ and $a_2(k)$, to find  $A_2(x)=A_1(x)$ for all $x$. 
% Note that the proof still holds if $b(k)$ vanishes at discrete values of $k$ (such as when $b(k)$ smoothly changes sign), because a lack  of knowledge of the value of  $a_i(k)$ at a few discrete points is not enough to affect $A_i(x)$.

This means that so long as the Fourier transform of all the probability distributions in Eq.~(\ref{Eq:evans-searles-step3}) are well-behaved, and that the Fourier transform of  $P\big(\Delta S^{\rm tot}_1;t_1,t_0\big)$ only vanishes at discrete points,
then
\begin{eqnarray}
Q\big(-\Delta S^{\rm tot}_{2\leftarrow1};t_2,t_1\big) &=& Q(\Delta S^{\rm tot}_{2\leftarrow1};t_2,t_1) \, e^{-\Delta S^{\rm tot}_{2\leftarrow1}}, \qquad
\end{eqnarray}
where $\Delta S^{\rm tot}_{2\leftarrow1}$ is the entropy change between time $t_1$ and time $t_2$,
given by Eq.~(\ref{Eq:S_1_to_2}).
This is the Evans-Searles steady-state fluctuation relation derived for a non-factorizable steady state 
in a non-Markovian system. The derivation holds for any situation where all initial system states decay to the same steady-state.

A careful reader will note that the proof is a little more general, it can also hold for a system with multiple steady-states (A, B, etc), 
so long as the initial factorized state with the system density matrix which corresponds to the reduced density matrix of steady-state A does indeed decay to steady-state A (and not to steady-state B).  This is plausible, since one would imagine that this initial state is the closest product state to steady-state A, but there may be systems that violate it.  Of course the proof does not apply to systems which do not decay to steady-states, such as those that decay to limit cycles.

\section{Approximate theories}
\label{Sect:approx}

This work connects fluctuation theorems to a microscopic symmetry of the system-reservoirs interactions,
going beyond Ref.~[\onlinecite{Campisi-review2011}].
This can be used to identify a family of approximations which are guaranteed to satisfy fluctuation theorems. These approximations must contains a trajectory $\overline\dtraj$ for every trajectory $\dtraj$, and individual transitions must satisfy local-detailed balance,
thereby satisfying  Eqs.~(\ref{Eq:D_0,1,2-TR}).
Then the above arguments apply, so Eq.~(\ref{Eq:central-obs}) is recovered,  which leads to all the usual fluctuation theorems, which means they will always obey the second law on average.

The first approximation is the Born approximation for weak system-reservoir coupling,
also called the Bloch-Redfield \cite{Bloch57,redfield} or sequential tunnelling approximation,\cite{Schoeller-review1997} see also Refs.~[\onlinecite{Nakajima58,Zwanzig60,Davies74,Davies76}] 
or various textbooks.\cite{Bellac-qm-book,Atom-Photon-Interactions-book,Blum-book}
This neglects trajectories where the system interacts with multiple reservoir modes at the same time, which is reasonable when the coupling is weak on the scale of the reservoir's memory time.
% The result is Fermi golden-rule rates for all system-reservoir interactions. 
The approximation has a trajectory $\overline\dtraj$ for every $\dtraj$, and individual transitions satisfy local-detailed balance,
which is enough to proof that it obeys all the usual fluctuation theorems.  
For strictly vanishing memory time (Markovian dynamics), this reduces to a Lindblad 
equation,\cite{lindblad,Davies74,whitney2008} for which 
a different proof of fluctuation theorems exists.\cite{Maxime+Alexia2016}  
However, our proof applies equally to  systems with short (but non-zero) memory times.

Next is the cotunnelling approximation, \cite{Schoeller-review1997} in which the system can interact with two reservoir modes at the same time.
This is a used in Coulomb-blockaded quantum dots, where it can dominate the transport in 
certain regimes.\cite{Schoeller-review1997} 
%There
%the simultaneous interaction with two reservoir modes (cotunnelling contribution) can dominate over the 
%interaction with one mode at a time (sequential tunnelling contribution) even at weak-coupling \cite{Schoeller-review1997}, if the charging energy is much larger than temperature or bias.
Since this approximation obeys the conditions discussed above,  this constitutes a proof that
the cotunnelling approximation obeys all the usual fluctuation theorems. 
Similarly, 
by allowing up to $n$ simultaneous interactions with reservoir modes (for different $n$),
one gets a family of approximations which all obey the fluctuation theorems.

\section{Conclusions}
This work uses a real-time diagrammatic theory on the Keldysh contour to develop the {\it quantum stochastic thermodynamics} of 
arbitrary systems coupled to ideal reservoirs.  It shows that energy conservation ensures that the system obeys the first law of thermodynamics on average.
Then, by finding the symmetry between trajectories on the Keldysh contour in Eq.~(\ref{Eq:central-obs}), 
it shows that the integral fluctuation theorem, Eq.~(\ref{Eq:non-equil_partition_identity}), holds for all non-Markovian system dynamics,
including non-factorized initial conditions, so these dynamics obey the second law on average.  
It gives other fluctuation theorems, such as Jarzynski or Crooks, in the right conditions.
Similarly, a non-factorized steady-state obeys the Evans-Searles fluctuation relation,\cite{Evans-Searles1994} 
if the Hamiltonian in Eq.~(\ref{Eq:H_tot}) is invariant under time-reversal.

The most obvious practical consequence of these results is that they prove that no quantum machine (Markovian or non-Markovian)
will ever exceed Carnot efficiency on average.  

A family of approximations  is identified which satisfies Eq.~(\ref{Eq:central-obs}), and so fulfill the fluctuation theorems.
This provides a powerful tool to analyse nanoscale energy-harvesting and refrigeration
beyond weak-coupling. 

%=======================================================

\begin{acknowledgments}
In fond memory of Maxime Clusel, whose ideas on quantum fluctuation theorems stimulated this work.  
I thank A.\ Auffeves, D.\ Basko, M.\ Campisi, A.\ Crepieux, C.\ Elouard,  M.\ Esposito, E.~Jussiau and F.\ Michelini 
for useful comments or discussions.
I acknowledges the financial support of the COST Action MP1209 ``Thermodynamics in the quantum regime'', the CNRS PEPS grant ``ICARE'', and the French National Research Agency's ``Investissement d'avenir'' program (ANR-15-IDEX-02) via the Universit\'e Grenoble Alpes QuEnG project.
\end{acknowledgments}

%=======================================================
%=======================================================

\appendix 
%====================================

\section{Reminder on time-reversal in quantum mechanics}
\label{Append:TR}

Here we recall the results that we will need related to time-reversal in quantum mechanics, which can be found in Messiah's famous textbook.\cite{Messiah-chapt15}
Firstly, the time-inversion of a quantum state $|i\rangle$ is defined as
\begin{eqnarray}
| \overline{\imath} \rangle &=& \hat{\Theta}^\dagger \, |i \rangle
\label{Eq:time-reverse-state-i}
\end{eqnarray}
where $\hat{\Theta}^\dagger$ is the time-inversion operator.
In the absence of spins, time-inversion of a wavefunction is just taking its complex conjugate;
thus $\hat{\Theta}^\dagger = \hat{\Theta}_0^\dagger$, where $\hat{\Theta}_0^\dagger$ is the complex-conjugation operator.
To understand the role of $\hat{\Theta}_0^\dagger$ for a single particle problem, one notes that position states are invariant under time-inversion, and so if one writes the system wavefunction $|i\rangle$ as a 
vector of position states, then $\hat{\Theta}_0^\dagger$ is the operator which takes the complex conjugate of all elements of the vector. For a many body problem, the same is also true, if one writes the system state as a vector of many-body position states (with a position for each particle).
Defining $\hat{\Theta}_0$ such that $\hat{\Theta}_0 \hat{\Theta}_0^\dagger = 1$, 
one has
$\hat{\Theta}_0^\dagger \,{\cal X}\, \hat{\Theta}_0 = {\cal X}^*$ 
for any matrix ${\cal X}$ written in a basis of many-body position states.

In the presence of spin-halves, the time-inversion operator also flips the spins about the y-axis, so 
\begin{eqnarray}
\hat{\Theta}^\dagger &=& -\rmi \sigma_y \,\hat{\Theta}^\dagger_0  \ .
%= \left(\begin{array}{cc} 0 & -1 \\ 1 & 0 \end{array}\right) \hat{\Theta}_0^\dagger 
\end{eqnarray}
The time-inversions of a position operator, $\hat{x}$, a momentum operator, $\hat{p} = -\rmi h\rmd / \rmd x$,
and a Paulli spin operator $\hat{\sigma}_\alpha$ are
\begin{subequations}
\begin{eqnarray}
\overline{\hat{x}}&=& \hat{\Theta}^\dagger \,\hat{x} \,\hat{\Theta} \ =\ \hat{x},
\\
\overline{\hat{p}}&=& \hat{\Theta}^\dagger \,\hat{p} \,\hat{\Theta}\ =\ -\hat{p},
\\
\overline{\hat{\sigma}}_\alpha &=& \hat{\Theta}^\dagger\, \hat{\sigma}_\alpha \,\hat{\Theta}\ =\ -\hat{\sigma}_\alpha,
\end{eqnarray}
\end{subequations}
The time reverse of a Hamiltonian in the time-window $t_0$ to $t$ as sketched in Fig.~\ref{Fig:a1} is
\begin{eqnarray}
\overline{\hat{H}\big(B,\sigma_\alpha,\tau \big) }
&=& \hat{\Theta}^\dagger\, \hat{H} \big(B,\hat{\sigma}_\alpha,t_0+t-\tau \big) \,\hat{\Theta}  
\nonumber \\
&=& \hat{H} \big(-B,-\hat{\sigma}_\alpha,t_0+t-\tau \big) \, ,
\label{Eq:H-t-reverse}
\end{eqnarray}
where the dependence of ${\cal H}$ on external fields, $B$, 
and Paulli spin-matrices, $\sigma_\alpha$, is explicitly shown to recall how they transform under time-reversal.
The evolution operator from time $t_0$ to time $\tau$  under such a time-dependent Hamiltonian (the solid part of the curve in Fig~\ref{Fig:a1}a)
is given by the usual time-ordered integral 
\begin{eqnarray}
\hat{U}(\tau;t_0) \ =\ {\cal T} \exp\left[-\rmi \int_{t_0}^\tau \rmd \tau' \hat{H}(\tau') \right]
\end{eqnarray}
where ${\cal T}$ is the time-ordering operator.
Similarly, the evolution operator  from time $\tau$ to time $t$  (the dashed part of the curve in Fig~\ref{Fig:a1}a) is
\begin{eqnarray}
\hat{U}(t;\tau) \ =\ {\cal T} \exp\left[-\rmi \int_\tau^t \rmd \tau' \hat{H}(\tau') \right]
\end{eqnarray}
If one now compares this to the evolution operator from time $t_0$ to time $\overline{\tau}$ in the system with the time-reversed Hamiltonian  (the solid part of the curve in Fig~\ref{Fig:a1}b)
\begin{eqnarray}
\overline{\hat{U}(t;\overline{\tau})}\ =\  \ {\cal T} \exp\left[-\rmi \int_{t_0}^{\overline{\tau}} \rmd \tau' \ \overline{\hat{H}(\tau')} \right] 
\end{eqnarray}
where one should recall that $\overline{\tau}=t_0+t-\tau$
Then it is straight-forward to show that
\begin{eqnarray}
\overline{{\hat U}(\overline{\tau};t_0)}
&=& \hat{\Theta}^\dagger \ \hat{U}^\dagger(t;\tau) \  \hat{\Theta} 
\label{Eq:HatU_and_Time-reverse}
\end{eqnarray}

%============================================
%============================================
%============================================


\begin{thebibliography}{1}
\bibitem{Footnote:units}
We take entropy in units of $\kB$,
temperature in units of energy, and time in units of $1/$(energy); so  $\hbar=\kB=1$.
\bibitem{footnote:error}
\dRed{
This condition was overlooked in the published version of this article,
due to stupid confusion between $\Delta S_{\rm res}(\dtraj)$ and $\Delta S_{\rm tot}(\dtraj)$. To be clear, one always has 
$\overline{\Delta S_{\rm res}}(\overline{\dtraj}) = -\Delta S_{\rm res}(\dtraj)$, but $S_{\rm tot}(\dtraj)$ also depends on the system's entropy change, which depends on the initial and final system states.
In general, to ensure that 
$\overline{\Delta S_{\rm tot}}(\overline{\dtraj}) = -\Delta S_{\rm tot}(\dtraj)$ for all $\dtraj$, one requires that $\rho_{\rm sys}^{\rm (f2)}=\rho_{\rm sys}^{\rm (i)}$.}
\end{thebibliography}
\end{document}